\documentclass[onecolumn,letter]{IEEEtran}
\usepackage{amsmath}
\usepackage{latexsym}
\usepackage{amssymb}
\usepackage{bm}
\usepackage[dvips]{graphicx}

\newtheorem{thm}    {Theorem}
\newtheorem{lem}     {Lemma}
\newtheorem{rem}     {Remark}

\newcommand{\specW}{\bm{W}}
\newcommand{\specWt}{\bm{W^{\times}}}
\newcommand{\specP}{\bm{P}}
\newcommand{\specPd}{\bm{P'}}
\newcommand{\specPt}{\bm{P^{\times}}}
\newcommand{\specQ}{\bm{Q}}
\newcommand{\specc}{\bm{c}}

\newcommand{\defeq}{\stackrel{\rm def}{=}}
\def\real{\mathbb R}
\def\rP{{\rm P}}
\def\dm{\mathop{\rm DM}}
\def\am{\mathop{\rm AM}}

\def\supp{\mathop{\rm supp}}
\def\argmax{\mathop{\rm argmax}}
\def\argmin{\mathop{\rm argmin}}
\def\plimsup{\hbox{p-}\limsup}
\def\pliminf{\hbox{p-}\liminf}
\def\plim{\hbox{p-}\lim}
\def\epliminf{\epsilon\hbox{-p-}\liminf}
\def\eepliminf{(1-\epsilon)\hbox{-p-}\liminf}

\def\ep{{\rm ep}}
\def\erf{G}

\def\Label#1{\label{#1}\ [\ #1\ ]\ }
\def\Label{\label}

\begin{document}
\title
{Information Spectrum Approach to\\ Second-Order Coding Rate in Channel Coding}
\author{
Masahito Hayashi
\thanks{
M. Hayashi is with Graduate School of Information Sciences, Tohoku University, Aoba-ku, Sendai, 980-8579, Japan
(e-mail: hayashi@math.is.tohoku.ac.jp)
}}
\date{}
\maketitle

\begin{abstract}
Second-order coding rate of channel coding
is discussed for general sequence of channels.
The optimum second-order transmission rate with a constant error constraint $\epsilon$
is obtained by using the information spectrum method.
We apply this result to 
the discrete memoryless case,
the discrete memoryless case with a cost constraint,
the additive Markovian case,
and 
the Gaussian channel case with an energy constraint.
We also clarify that 
the Gallager bound does not give the optimum evaluation in the second-order coding rate.
\end{abstract}

\section{Introduction}\Label{s1}
\PARstart{B}{ased} on the channel coding theorem, there exists a sequence of codes for the given channel $W$ such that the average error probability goes to $0$ when the transmission rate $R$ is less than $C^{\dm}_W$. That is, if the number $n$ of applications of the channel $W$ is sufficiently large, the average error probability of a good code goes to $0$. In order to evaluate the average error probability with finite $n$, we often use the exponential rate of decrease, which depends on the transmission rate $R$. However, such an exponential evaluation ignores the constant factor. Therefore, it is not clear whether exponential evaluation provides a good evaluation for the average error probability when the transmission rate $R$ is close to the capacity. In fact, many researchers believe that, out of the known evaluations,
 the Gallager bound \cite{Gal} gives the best upper bound of average error probability in the channel coding when the transmission rate is greater than the critical rate. 
This is because the Gallager bound provides the optimal exponential rate of decrease. 
In order to clarify this point, we focus on the second-order coding rate in channel coding, in which, we describe the transmission length by $C^{\dm}_W n +R_2 \sqrt{n}$. 
From a practical viewpoint, when the coding length is close to $C^{\dm}_W n$, the second-order coding rate gives a better evaluation of average error probability than the first-order coding rate. 
In fact, the second error coding rate has been applied for evaluation of the average error probability of random coding concerning the phase basis, which is essential to the security of quantum key distribution\cite{H-QKD}. Therefore, it is appropriate to treat the second-order coding rate from the applied viewpoint as well as the theoretical viewpoint.
In the case of the discrete memoryless case,
Strassen \cite{strassen} derived the optimum rate $R_2$ for an arbitrary average error probability $0 < \epsilon< 1$ using the Gaussian distribution.
In the present paper, we extend his result to more general cases, i.e., the discrete memoryless case with cost constraint, the Gaussian additive noise case with the energy constraint, and the additive Markovian case.
Further, our proof for the discrete memoryless case is much simpler than the original one.
Indeed, since his proof is not so simple and his paper is written in German, it is quite difficult to follow his proof.

In the present paper, in order to treat this problem from a unified viewpoint,
we employ the method of information spectrum, which was initiated by Han-Verd\'{u} \cite{Han-Verdu}, and was mainly formulated by Han\cite{Han1}. 
The second-order coding rate is closely related to the method of information spectrum
because Hayashi\cite{H-sec} treated this problem of fixed-length source coding and intrinsic randomness using the method of information spectrum.
Hayashi\cite{H-sec} discussed the error probability when the compressed size is $H(P) n + a \sqrt{n}$, where $n$ is the size of input system and $H(P)$ is the entropy of the distribution $P$ of the input system.
In the method of information spectrum, we treat the general asymptotic formula, which gives the relationship between the asymptotic optimal performance and the normalized logarithm of the likelihood of the probability distribution.
In order to treat a special case, we apply the general asymptotic formula to the respective information source and calculate the asymptotic stochastic behavior of the normalized logarithm of the likelihood. That is, in the information spectrum method, we have two steps, deriving the general asymptotic formula and applying the general asymptotic formula. With respect to fixed-length source coding and intrinsic randomness, the same relation holds concerning the general asymptotic formula in the second-order coding rate. However, there is a difference concerning the application of the general asymptotic formula to the independent and identical distributions. That is, while the normalized logarithm of the likelihood approaches the entropy $H(P)$ in the probability in the first-order coding rate, the stochastic behavior is asymptotically described by the Gaussian distribution in the first-order coding rate. In other words, in the second step, the first-order coding rate corresponds to the law of large numbers, and the second-order coding rate corresponds to the central limit theorem. 

In the present paper, we treat the channel coding in the second-order coding rate, i.e., the case in which the transmission length is $C^{\dm}_W n + a \sqrt{n}$. Similar to the above-mentioned case, we employ the method of information spectrum. That is, we treat the general channel, which is the general sequence $\{W^n(y|x)\}$ of probability distributions without structure. As shown by Verd\'{u}-Han \cite{Verdu-Han}, this method enables us to characterize the asymptotic performance with only the random variable $\frac{1}{n}\log \frac{W^n(y|x)}{W^n_{P^n}(y)}$ (the normalized logarithm of the likelihood ratio between the conditional distribution and the non-conditional distribution) without any further assumption, where $W^n_{P^n}(y)\defeq\sum_{x}P^n(x)W^n(y|x)$. Concerning this general asymptotic formula, if we can suitably formulate theorems in the second-order coding rate and establish an appropriate relationship between the first-order coding rate and the second-order coding rate, we can easily extend proofs concerning the first-order coding rate to those of the second-order coding rate. Therefore, there is no serious difficulty in establishing the general asymptotic formula in the second-order coding rate. In order to clarify this point, we present proofs of some relevant theorems in the first-order coding rate, even though they are known.

In order to treat the special cases,
it is sufficient to apply the general asymptotic formula, i.e., to calculate the asymptotic behavior 
of the random variable $\frac{1}{n}\log \frac{W^n(y|x)}{W^n_{P^n}(y)}$.
The additive Markovian case can be 
treated in the same way as fixed-length source coding and intrinsic randomness.
However, other special cases have another difficulties,
which do not appear in fixed-length source coding or intrinsic randomness.
The first difficulty is
the optimization concerning the input distribution in the converse part of the channel coding. 
This problem commonly appears among the
three cases, i.e., the discrete memoryless case,
the discrete memoryless case with cost constraint,
and the Gaussian additive noise case with the energy constraint.
In the discrete memoryless case, the second-order coding rate corresponds to simple application of the central limit theorem, while the first-order coding rate corresponds to the law of large numbers. 
Hence, the performance in second-order coding rate is characterized by the variance of the logarithmic likelihood ratio, and the direct part can be easily obtained in this case.
This relationship is summarized in Fig. \ref{pic3}. 

\begin{figure}[htbp]
\begin{center}
\scalebox{1.0}{\includegraphics[scale=0.3]{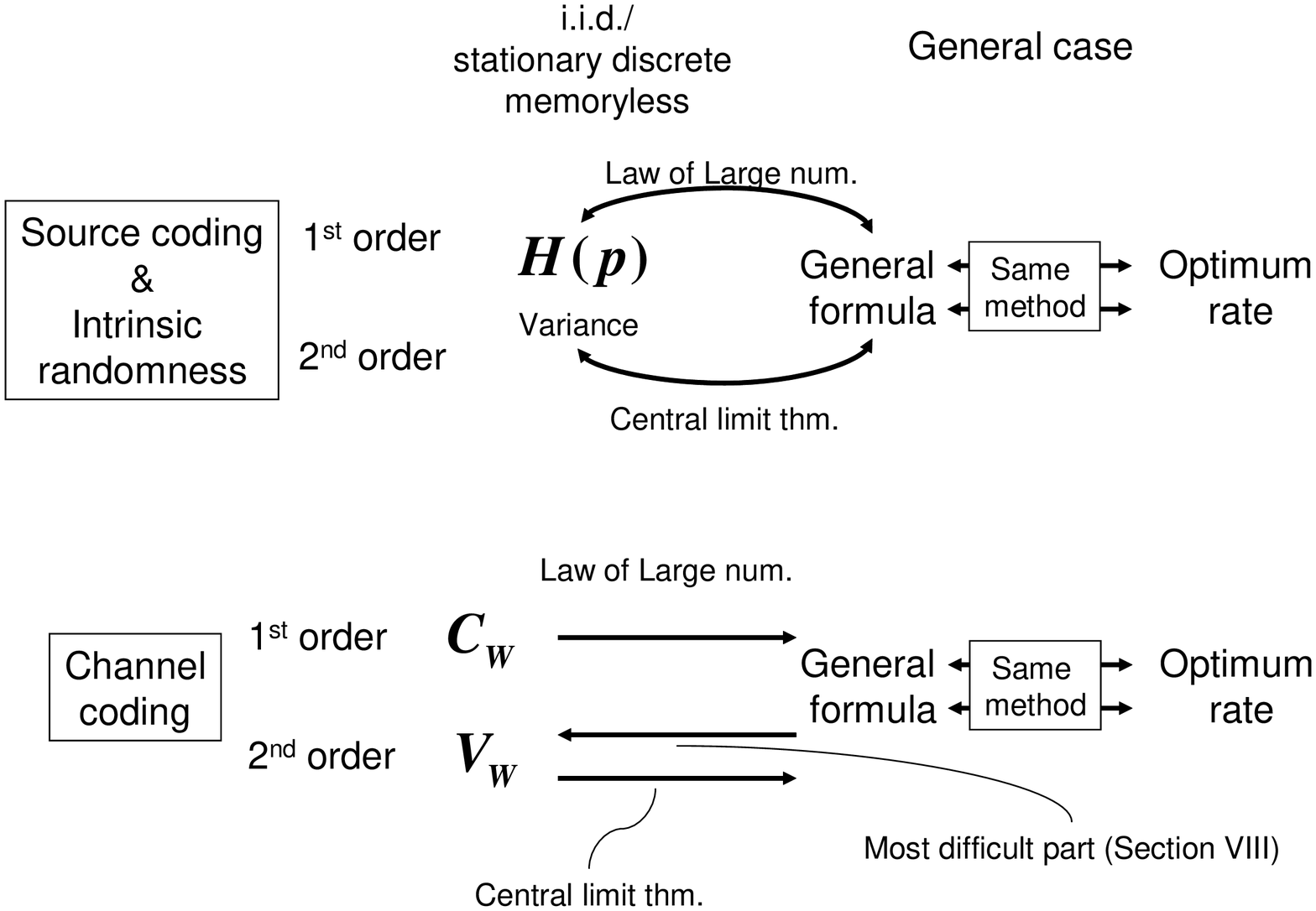}}
\end{center}
\caption{Relationship between the present result and fixed-length source coding/intrinsic randomness.
The $\to$ arrow describes the direct part, and 
the $\leftarrow$ arrow describes the converse part.}
\Label{pic3}
\end{figure}%

However, there is another difficulty in the direct part for the discrete memoryless case with cost constraint and the Gaussian additive noise case with the energy constraint.
In these cases, all of the encoded signals has to satisfy cost constraint.
This kind of difficulty does not appear in the case of first-order coding rate of both of the discrete memoryless case with cost constraint and the Gaussian additive noise case with the energy constraint.
This is because it is sufficient to 
construct the code whose average error probability goes to zero in the case of the first-order coding rate while it is required to construct the code whose average error probability goes to a given thereshold $\epsilon$ in the case of the second-order coding rate.
When we find a code satisfying the following; its average error probability goes to zero and its average cost is less than the constraint.
Then, there exists a subcode satisfying the following; its average error probability goes to zero and the costs of all encoded signals are less than the constraint.
However, the same method cannot be applied when 
we find a code 
satisfying the following; its average error probability goes to $\epsilon$ and its average cost is less than the constraint.
In the present paper, we directly construct a code, in which
the costs of all encoded signals are less than the constraint.

Here, we describe the meaning of the second-order coding rate.
When the transmission length is described by $nC^{\dm}_W +\sqrt{n} R_2$, as shown in Subsection \ref{s9a}, the optimal error can be approximately attained by random coding. Since it seems that random coding cannot be realized, 
our evaluation seems to be related to only the theoretical best performance.
However, in the quantum key distribution, it can be realized concerning the phase bases \cite{WMT,H-QKD}.
In such a setting, the coding length is on the order of 10,000 or 100,000\cite{HHHTT}.
In the quantum key distribution, Hayashi \cite{H-QKD} has applied the second-order coding rate to evaluate the phase error probability, which is directly linked to the security of the final key.

The remainder of the present paper is organized as follows.
In Section \ref{s2}, we revisit the second order coding rate 
in the stationary discrete memoryless case, and dicuss 
the second order coding rate in the stationary discrete memoryless case with cost constraint.
In Section \ref{s3}, 
the Markovian additive channel is treated.
In Section \ref{s4}, 
the Gaussian additive noise case with the energy constraint
is considered.
These results are shown in the Section \ref{s10} by employing the method of information specturm. 
In the present result,  
the performance of information transmission is discussed 
in terms of second-order coding rate using 
two important quantities $V_W^+$ and $V_W^-$ 
instead of the capacity in the case of discrete memoryless case.
In other cases, similar quantities play the same role.

In Section \ref{s5}, we compare our evaluation with 
the Gallager bound \cite{Gal} in the second-order setting.
In Section \ref{s6}, the properties of $V_W^+$ and $V_W^-$ are discussed.
In Subsection \ref{s6a}, we discuss a typical example such that 
$V_W^+$ is different from $V_W^-$.
In Subsection \ref{s6b}, the additivities concerning $V_W^+$ and $V_W^-$ are proved.
In Section \ref{s7}, the notations of the information spectrum are explained.
In Section \ref{s8}, the performance of the information transmission is discussed 
in terms of the second-order coding rate 
using the information spectrum in the general case.
That is, we present general formulas for the second-order coding rate.
In Section \ref{s9}, the theorem presented in the previous section is proved.
In Section \ref{s10}, using general formulas for the second-order coding rate,
we demonstrate our proof of the second order coding rate in the stationary discrete memoryless case
using our general result concerning the second order coding rate. 
In this proof, the direct part is immediate.
The converse part is the most difficult considered herein 
because we must treat the information spectrum for the general input distributions
in the sense of the second-order coding rate.

\section{Second order coding rate in stationary discrete memoryless channels}\Label{s2}
As the most typical case, we revisit the second-order coding rate of stationary discrete memoryless channels, in which, we use an $n$-multiple application of the discrete channel $W(y|x)$, which transmits the information from the input system ${\cal X}$
to the output system ${\cal Y}$.
That is, the channel considered here is given as the stationary discrete memoryless channel 
$W^{\times n}(y|x)\defeq\prod_{i=1}^n W(y_i|x_i)$.
Note that, in the present paper,
$P\times P'$ ($W\times W'$) denotes the product of two distributions $P$ and $P'$
(two channels $W$ and $W'$), and
$P^{\times n}$ ($W^{\times n}$) denotes 
the product of $n$ uses of the distribution $P$ (the channel $W$), i.e.,
the $n$-th independent and identical distribution (i.i.d.) of $P$ 
(the $n$-th stationary memoryless channel of $W$). 
In this case, when the transmission rate is less than the capacity $C_W^{\dm}$,
the average error probability goes to $0$ exponentially, if we use 
a suitable encoder and the maximum likelihood decoder.

Let $N$ be the size of the transmitted information. 
The encoder is a map $\phi$ from $\{1, \ldots, N\}$ to 
${\cal X}^n$, and the decoder is 
given by the set of subsets $\{{\cal D}_i\}_{i=1}^N$ of ${\cal Y}^n$, where
${\cal D}_i$ corresponds to the decoding region of $i \in \{1, \ldots, N\}$.
Then, the code is given by the triple $(N,\phi,\{{\cal D}_i\}_{i=1}^N)$ 
and is denoted by $\Phi$.
The average error probability $P_{e,W^{\times n}}(\Phi )$ is described as
\begin{align*}
P_{e,W^{\times n}}(\Phi )\defeq
\frac{1}{N_n} \sum_{i=1}^{N_n} (1- W^{\times n}_{\phi(i)}({\cal D}_i )),
\end{align*}
where $W_x(y)\defeq W(y|x)$.
For simplicity, the size $N_n$ is denoted by $|\Phi|$.
The performance of the code $\Phi$ is given by the pair of $P_e(\Phi )$ and $|\Phi|$.
As stated by the channel coding theorem \cite{Shannon1}, the capacity is given by 
\begin{align*}
C_W^{\dm}=\max_{P} I(P,W)
=\min_{Q} \max_{x}  D(W_x\|Q),
\end{align*}
where $Q$ is the output distribution, and
\begin{align*}
W_P(y)&\defeq \sum_{x} P(x)W(y|x) \\
I(P,W)&\defeq \sum_{x} P(x) D(W_x\|W_P)\\
D(P\|P')&\defeq \sum_{x} P(x)\log \frac{P(x)}{P'(x)}.
\end{align*}
Thus, $Q_M\defeq \argmin_{Q}\max_{x} D(W_x\|Q)$ satisfies
\begin{align}
D(W_x\|Q_M)\le C^{\dm}_W \Label{12-5-9}.
\end{align}
Throughout the present paper, we choose the base of the logarithm to be $e$.

Although the above channel coding theorem concerns only the first-order coding rate of the transmission length $\log N_n$, 
our main focus is the analysis of the second-order coding rate.  
When the transmission length $\log N_n$ asymptotically behaves as $nC^{\dm}_W + a \sqrt{n}$,
the optimal average error is given as follows:
\begin{align}
C_p^{\dm}(a,C_W^{\dm}|W)\defeq 
\inf_{\{\Phi_n\}_{n=1}^{\infty}
}
\left\{\left.
\limsup_{n \to \infty} P_{e,W^{\times n}}(\Phi_n )
\right|
\liminf_{n \to \infty}
\frac{1}{\sqrt{n}}(\log |\Phi_n|- nC^{\dm}_W) \ge a
\right\}. \Label{11-30-3}
\end{align}
Fixing the average error probability,
we obtain the following quantity:
\begin{align}
C^{\dm}(\epsilon, C_W^{\dm}|W)\defeq 
\sup_{\{\Phi_n\}_{n=1}^{\infty}
}
\left\{\left.
\liminf_{n \to \infty}
\frac{1}{\sqrt{n}}(\log |\Phi_n|- nC^{\dm}_W) 
\right|
\limsup_{n \to \infty} P_{e,W^{\times n}}(\Phi_n ) \le \epsilon
\right\}.\Label{11-30-4}
\end{align}
We refer to this value the optimum second-order transmission rate with the error probability $\epsilon$.
In order to treat the second-order coding rate,
we need the distribution function $\erf$ for the standard Gaussian distribution
(with expectation $0$ and variance $1$),
which is defined by
\begin{align*}
\erf(x) &\defeq \int_{-\infty}^{x}
\frac{1}{\sqrt{2\pi}}
e^{- x^2/2}\,d x.
\end{align*}
In this problem, 
the quantity $V_{P,W}$:
\begin{align*}
V_{P,W} &\defeq \sum_{x}P(x)\sum_y W_x(y)
\left(\log \frac{W_x(y)}{W_P(y)} -D (W_x\|W_P)\right)^2
\end{align*}
plays an important role.
By using these quantities, 
$C_p^{\dm}(a,C_W^{\dm}|W)$ and $C^{\dm}(\epsilon,C_W^{\dm}|W)$ are calculated 
in the stationary discrete memoryless case as follows
\begin{thm}(Strassen\cite{strassen})\Label{thm1}
When the cardinality $|{\cal X}|$ is finite and 
$P_M\defeq\argmax_P I(P,W)$ exists uniquely, then 
\begin{align}
C_p^{\dm}(a,C_W^{\dm}|W)&= \erf(a/\sqrt{V_{P_M,W}})\Label{12-5-11}\\
C^{\dm}(\epsilon,C_W^{\dm}|W) &= \sqrt{V_{P_M,W}} \erf^{-1}(\epsilon). \Label{12-5-12}
\end{align}
\end{thm}
When $\{W_x\}$ is linearly independent by regarding distributions as positive vectors,
the map $P \mapsto W_P$ is a one-to-one map.
Then, $P_M\defeq\argmax_P I(P,W)$ exists uniquely.
However, when $\{W_x\}$ is not linearly independent,
$\argmax_P I(P,W)$ is not necessarily unique.
In order to treat such a case,
we introduce two quantities $V_{W}^+$ and $V_{W}^-$ and
two distributions $P_{M+}$ and $P_{M-}$:
\begin{align*}
V_W^+ &\defeq \max_{P\in {\cal V}} V_{P,W} \\
V_W^- &\defeq \min_{P\in {\cal V}} V_{P,W} \\
P_{M+}&\defeq \argmax_{P\in {\cal V}} V_{P,W} \\
P_{M-}&\defeq \argmin_{P\in {\cal V}} V_{P,W},
\end{align*}
where ${\cal V}\defeq\{P|I(P,W)=C^{\dm}_W\}$.
In order to treat such a case, Theorem \ref{thm1} is generalized as follows:
\begin{thm}(Strassen\cite{strassen})\Label{thm2}
When the cardinality $|{\cal X}|$ is finite and the set ${\cal V}$ has multiple elements,  
(\ref{12-5-11}) and (\ref{12-5-12}) are generalized as
\begin{align*}
C^{\dm}_p(a,C^{\dm}_W|W)&= 
\left\{ 
\begin{array}{cc}
\erf(a/\sqrt{V_W^+}) & a \ge 0 \\
\erf(a/\sqrt{V_W^-}) & a < 0
\end{array}
\right. \\
C^{\dm}(\epsilon,C^{\dm}_W|W) &= 
\left\{ 
\begin{array}{cc}
\sqrt{V_W^+} \erf^{-1}(\epsilon) & \epsilon \ge 1/2\\
\sqrt{V_W^-} \erf^{-1}(\epsilon) & \epsilon < 1/2.
\end{array}
\right.
\end{align*}
More precisely, the direct part
\begin{align}
C^{\dm}_p(a,C^{\dm}_W|W)& \le
\left\{ 
\begin{array}{cc}
\erf(a/\sqrt{V_W^+}) & a \ge 0 \\
\erf(a/\sqrt{V_W^-}) & a < 0
\end{array}
\right.\Label{8-10-1}\\
C^{\dm}(\epsilon,C^{\dm}_W|W) & \ge
\left\{ 
\begin{array}{cc}
\sqrt{V_W^+} \erf^{-1}(\epsilon) & \epsilon \ge 1/2\\
\sqrt{V_W^-} \erf^{-1}(\epsilon) & \epsilon < 1/2.
\end{array}
\right.\Label{8-10-2}
\end{align}
hold without any assumption,
and 
the converse part
\begin{align*}
C^{\dm}_p(a,C^{\dm}_W|W)& \ge
\left\{ 
\begin{array}{cc}
\erf(a/\sqrt{V_W^+}) & a \ge 0 \\
\erf(a/\sqrt{V_W^-}) & a < 0
\end{array}
\right.\\
C^{\dm}(\epsilon,C^{\dm}_W|W) &\le
\left\{ 
\begin{array}{cc}
\sqrt{V_W^+} \erf^{-1}(\epsilon) & \epsilon \ge 1/2\\
\sqrt{V_W^-} \erf^{-1}(\epsilon) & \epsilon < 1/2.
\end{array}
\right.
\end{align*}
hold with the assumption $|{\cal X}|<\infty$.
\end{thm}

Next, 
consider the cost function $c : {\cal X} \mapsto \real$.
In this case, we often assume that
all encoded alphabets $\phi(i)$ of the code $\Phi_n$ belongs to the set 
\begin{align*}
{\cal X}^n_{c,K}\defeq 
\left \{x\in {\cal X}^n\left |
\sum_{i=1}^n c(x_i) \le K \right.\right\}.
\end{align*}
The maximum coding rate with the above condition is called the capacity with the cost constraint,
and is given by \cite{CK}
\begin{align*}
C_{W,c,K}^{DM}
= 
\max_{P:{\rm E}_P c(x)\le K} I(P,W)
=\min_{Q}
\max_{P:{\rm E}_P c(x)\le K} J(P,Q,W),  
\end{align*}
where 
\begin{align*}
J(P,Q,W)\defeq \sum_{x\in {\cal X}} P(x) D(W_x\|Q).
\end{align*}
In the same way to (\ref{11-30-3}) and (\ref{11-30-4}),
we define the following values with the cost constraint:
\begin{align}
C_p^{\dm}(a,C_W^{\dm}|W,c,K)
\defeq &
\inf_{\{\Phi_n\}_{n=1}^{\infty}
}
\left\{\left.
\limsup_{n \to \infty} P_{e,W^{\times n}}(\Phi_n )
\right|
\liminf_{n \to \infty}
\frac{1}{\sqrt{n}}(\log |\Phi_n|- nC^{\dm}_W) \ge a,
\supp(\Phi_n) \subset {\cal X}^n_{c,K}
\right\}. \Label{11-30-3b} \\
C^{\dm}(\epsilon, C_W^{\dm}|W,c,K)
\defeq &
\sup_{\{\Phi_n\}_{n=1}^{\infty}
}
\left\{\left.
\liminf_{n \to \infty}
\frac{1}{\sqrt{n}}(\log |\Phi_n|- nC^{\dm}_W) 
\right|
\limsup_{n \to \infty} P_{e,W^{\times n}}(\Phi_n ) \le \epsilon,
\supp(\Phi_n) \subset {\cal X}^n_{c,K}
\right\},\Label{11-30-4b}
\end{align}
where $\supp(\Phi_n)$ expresses the set $\{\phi(1), \ldots, \phi(N)\}$ for a code 
$\Phi=(N,\phi,\{{\cal D}_i\}_{i=1}^N)$.
We introduce two quantities $V_{W,c,K}^+$ and $V_{W,c,K}^-$ and
two distributions $P_{M+,c,K}$ and $P_{M-,c,K}$:
\begin{align*}
V_{W,c,K}^+ &\defeq \max_{P\in {\cal V}_{c,K}} V_{P,W} \\
V_{W,c,K}^- &\defeq \min_{P\in {\cal V}_{c,K}} V_{P,W} \\
P_{M+,c,K}&\defeq \argmax_{P\in {\cal V}_{c,K}} V_{P,W} \\
P_{M-,c,K}&\defeq \argmin_{P\in {\cal V}_{c,K}} V_{P,W},
\end{align*}
where ${\cal V}_{c,K}\defeq\{P|I(P,W)=C^{\dm}_{W,c,K}, {\rm E}_P c(x)\le K \}$.
\begin{thm}\Label{thm2b}
When the cardinality $|{\cal X}|$ is finite 
\begin{align*}
C^{\dm}_p(a,C^{\dm}_{W,c,K}|W,c,K)&= 
\left\{ 
\begin{array}{cc}
\erf(a/\sqrt{V_{W,c,K}^+}) & a \ge 0 \\
\erf(a/\sqrt{V_{W,c,K}^-}) & a < 0
\end{array}
\right.\\
C^{\dm}(\epsilon,C^{\dm}_{W,c,K}|W,c,K) &= 
\left\{ 
\begin{array}{cc}
\sqrt{V_{W,c,K}^+} \erf^{-1}(\epsilon) & \epsilon \ge 1/2\\
\sqrt{V_{W,c,K}^-} \erf^{-1}(\epsilon) & \epsilon < 1/2.
\end{array}
\right.
\end{align*}
More precisely, the direct part 
\begin{align}
C^{\dm}_p(a,C^{\dm}_{W,c,K}|W,c,K)&\le
\left\{ 
\begin{array}{cc}
\erf(a/\sqrt{V_{W,c,K}^+}) & a \ge 0 \\
\erf(a/\sqrt{V_{W,c,K}^-}) & a < 0
\end{array}
\right.\Label{8-10-3}\\
C^{\dm}(\epsilon,C^{\dm}_{W,c,K}|W,c,K) &\ge
\left\{ 
\begin{array}{cc}
\sqrt{V_{W,c,K}^+} \erf^{-1}(\epsilon) & \epsilon \ge 1/2\\
\sqrt{V_{W,c,K}^-} \erf^{-1}(\epsilon) & \epsilon < 1/2.
\end{array}
\right.\Label{8-10-4}
\end{align}
hold without any assumption,
and 
the converse part
\begin{align*}
C^{\dm}_p(a,C^{\dm}_{W,c,K}|W,c,K)&\ge
\left\{ 
\begin{array}{cc}
\erf(a/\sqrt{V_{W,c,K}^+}) & a \ge 0 \\
\erf(a/\sqrt{V_{W,c,K}^-}) & a < 0
\end{array}
\right.\\
C^{\dm}(\epsilon,C^{\dm}_{W,c,K}|W,c,K) &\le
\left\{ 
\begin{array}{cc}
\sqrt{V_{W,c,K}^+} \erf^{-1}(\epsilon) & \epsilon \ge 1/2\\
\sqrt{V_{W,c,K}^-} \erf^{-1}(\epsilon) & \epsilon < 1/2.
\end{array}
\right.
\end{align*}
hold with the assumption $|{\cal X}|<\infty$.
\end{thm}

\begin{rem}\rm
When the sets 
${\cal X}$ and ${\cal Y}$ are given as general probability spaces with general $\sigma$-fields
$\sigma({\cal X}_n)$ and $\sigma({\cal Y}_n)$,
the above formulation can be extended with the following definition.
The channel $W$ is given by the real-valued function from 
${\cal X}$ and $\sigma({\cal Y})$ satisfying the following conditions;
(i) For any $x \in {\cal X}$, $W^n$ is a probability measure on ${\cal Y}$,
(ii) For any $F \in \sigma({\cal Y})$, $W_{\cdot} (F)$ is a measurable function on ${\cal X}$.
$P$ take values in probability measures on ${\cal X}$.
Then, the summands   
$\sum_{x\in {\cal X}} P(x)$ and 
$\sum_{y\in {\cal Y}} W_x(y)$ are replaced by 
$\int_{{\cal X}} P( d x)$ and 
$\int_{{\cal Y}} W_x(d y)$, respectively.
For any distribution $Q$ on ${\cal Y}$,
the function 
$\frac{W_x(y)}{W_P(y)}$ is replaced by the inverse of Radon-Nikodym derivative 
$\frac{d W_P}{d W_x}(y)$ of $W_P$ with respect to $W_x$.
In this extension, the direct part (\ref{8-10-1}), (\ref{8-10-2}), (\ref{8-10-3}), and (\ref{8-10-4}) are valid.
\end{rem}

\section{Second order coding rate in additive Markovian channel}\Label{s3}
Next, we we focus on the additive Markovian channel, in which, 
we assume that the additive noise obeys the transition matrix $Q(y|x)$ on the set ${\cal X}=\{1, \ldots, d\}$.
Then, the channel $W(Q)^n(y|x)$ has the form $\prod_{i=1}^n Q(y_i-x_i|y_{i-1}-x_{i-1})$, where
$y_0-x_0$ is the initial state $s_0$ and the arithmetic is based on mod $d$.
For simplicity, we assume that 
the transition matrix $Q(y|x)$ is irreducible.
Then, the $n$-th marginal distribution
$Q^n(x_n):=\sum_{i_1,\ldots, i_n} \prod_{i=1}^n Q(x_i|x_{i-1})$ approaches the stationary distribution
$P_Q(x)$, which is given as the eigenvector of $Q(y|x)$ associated with the eigenvalue $1$\cite{DZ}.
When the conditional distribution $Q(y|x)$ is denoted by $Q_x(y)$,
the normalized entropy of the distribution $Q^n(\vec{x}_n):=\prod_{i=1}^n Q(x_i|x_{i-1})$ goes to 
$H(Q):=\sum_x P_Q(x) H(Q_x)$. 
Then, 
by defining the capacity $C_W^{\am}$ in the same way as $C_W^{\dm}$,
the channel capacity $C_W^{\am}$ is
calculated as
\begin{align}
C_Q^{\am}= \log d - H(Q).\Label{6-11-14}
\end{align}
Similar to 
$C_p^{\dm}(a,C_W^{\dm}|W)$ and $C^{\dm}(\epsilon,C_W^{\dm}|W)$,
the second order quantities $C_p^{\am}(a,C_Q^{\am}|W)$ and $C^{\am}(\epsilon,C_Q^{\am}|W)$
are defined for the additive Markovian case.
Then, the following theorem holds.
In this problem, 
the variance $V(Q)$:
\begin{align*}
&V(Q)\\
:=&
\sum_{y,x}Q(y|x)P_Q(x)
(-\log Q(y|x)-H(Q))^2 \\
& + 
2
\sum_{z,y,x}Q(z|y)Q(y|x)P_Q(x)
(-\log Q(z|y)-H(Q))
(-\log Q(y|x)-H(Q)).
\end{align*}
plays an important role.
By using these quantities, 
$C_p^{\am}(a,C_Q^{\am}|W)$ and $C^{\am}(\epsilon,C_Q^{\am}|W)$ are calculated 
in the additive Markovian case as follows
\begin{thm}\Label{thm11}
The relations
\begin{align*}
C_p^{\am}(a,C_Q^{\am}|W)&= G(a/\sqrt{V(Q)})\\
C^{\am}(\epsilon,C_Q^{\am}|W) &= \sqrt{V(Q)} G^{-1}(\epsilon). 
\end{align*}
hold.
\end{thm}


\section{Second order coding rate in Gaussian channel}\Label{s4}
In this section, we consider the case of additive Gaussian noise.
In this case, both of the input system and the output system are given by $\real$,
and 
the output distribution $W_x(y)$ is given by $\frac{1}{\sqrt{2\pi N}} e^{-\frac{(y-x)^2}{2 N}}$ 
for a given noise level $N$.
If there is no restriction for input signal, the capacity diverges.
Hence, it is natural to consider the cost constraint.
Consider the cost function $c(x)\defeq x^2$ and the maximum cost $S$.
Then, the maximum mutual information $\max_{P:{\rm E}_P x^2 \le S} I(P,W)$
is attained when $P$ is equal to $P_M(x)\defeq\frac{1}{\sqrt{2\pi S}} e^{-\frac{x^2}{2S}}$.
In this case, 
\begin{align}
D(W_x\|W_{P_M})=\frac{1}{2}\log (1+\frac{S}{N})+\frac{\frac{x^2}{N}-\frac{S}{N}}{2(1+\frac{S}{N})}.
\Label{8-8-1}
\end{align}
Then, the capacity is known to be \cite{Shannon1,Shannon2}
\begin{align*}
C^G_{N,S}= 
\max_{P:{\rm E}_P x^2 \le E} I(P,W)=
\frac{1}{2}\log (1+\frac{S}{N}).
\end{align*}
Since
\begin{align*}
\int_{-\infty}^\infty
\left(\log \frac{W_x(y)}{W_{P_M}(y)}- D(W_x\|W_{P_M}) \right)^2
W_x(y)dy =
\frac{\frac{S^2}{N^2} + 2 \frac{x^2}{N}}{2(1+\frac{S}{N})^2},
\end{align*}
$V_{P_M,W}$ is calculated as
\begin{align*}
V_{P_M,W}= 
\frac{\frac{S^2}{N^2} + 2 \frac{S}{N}}{2(1+\frac{S}{N})^2}.
\end{align*}
Since the cardinality of $\real$ is infinite,
the assumption of section \ref{s2} does not hold.
That is, we cannot apply Theorem \ref{thm2b}.
However, the following theorem holds.
\begin{thm}\Label{thm2c}
Define the quantities
$C^{G}_p(a,C^G_{N,S}|N,S)$
and
$C^{G}(\epsilon,C^G_{N,S}|N,S)$
in the same way as (\ref{11-30-3b}) and (\ref{11-30-4b}).
Then,
\begin{align*}
C^{G}_p(a,C^G_{N,S}|N,S)
&= 
\erf(a/\sqrt{V_{P_M,W}}) 
\\
C^{G}(\epsilon,C^G_{N,S}|N,S) &= 
\sqrt{V_{P_M,W}} \erf^{-1}(\epsilon).
\end{align*}
\end{thm}

\section{Comparison with the Gallager bound}\Label{s5}
At first glance, the Gallager bound \cite{Gal} seems to work well 
for evaluating the average error probability, even when the transmission length is close to $nC^{\dm}_W$.
This is because this bound gives the optimal exponential rate when the coding rate is greater than the critical rate.
In this section, we clarify whether the present evaluation or the Gallager bound \cite{Gal}
provides a better evaluation when the transmission length is close to $nC^{\dm}_W$.
For this analysis,
we describe the transmission length by $nC^{\dm}_W +\sqrt{n} R_2$.
Let us compare the present evaluation with the Gallager bound,
which is given by
\begin{align}
\min_{\Phi:|\Phi|\le e^{n R}}  P_{e,W^{\times n}}(\Phi)
\le \min_{P} \min_{0 \le s \le 1} e^{n (Rs +\psi_P(s))},\Label{Gallager}
\end{align}
where
\begin{align*}
\psi_P(s)\defeq \log 
\sum_y \left(
\sum_x P(x) W_x(y)^{\frac{1}{1+s}}
\right)^{1+s}.
\end{align*}
Since the present evaluation is essentially based on Verd\'{u}-Han's method\cite{Verdu-Han},
this comparison can be regarded as a comparison between 
Verd\'{u}-Han's evaluation and the Gallager bound.
Next, we substitute $nC^{\dm}_W + \sqrt{n} R_2$ into $nR$.
Then, 
\begin{align*}
\min_{0 \le s \le 1} e^{n (Rs +\psi_P(s))}
=  e^{ n\min_{0 \le s \le 1}(C^{\dm}_Ws + \frac{R_2}{\sqrt{n}} s+\psi_P(s))}.
\end{align*}
Taking the derivatives of $\psi_P(s)$, we obtain
\begin{align*}
\left. \frac{d \psi_P(s)}{d s}\right|_{s=0}&= -I(P,W)\\
\left. \frac{d^2 \psi_P(s)}{d s^2}\right|_{s=0}&= V_{P,W}.
\end{align*}
When $C^{\dm}_W=I(P,W)$,
\begin{align*}
& C^{\dm}_W s + \frac{R_2}{\sqrt{n}} s+\psi_P(s)
\cong C^{\dm}_W s + \frac{R_2}{\sqrt{n}} s -I(P,W)s +\frac{V_{P,W}}{2}s^2\\
=& \frac{R_2}{\sqrt{n}} s +\frac{V_{P,W}}{2}s^2
= \frac{V_{P,W}}{2}(s+ \frac{R_2}{\sqrt{n}V_{P,W}})^2
- \frac{R_2^2}{2nV_{P,W}}.
\end{align*}
Therefore, as is rigorously shown in Appendix, when $R_2 <0$,
\begin{align}
\lim_{n \to \infty}
n\min_{0 \le s \le 1}\left(C^{\dm}_Ws + \frac{R_2}{\sqrt{n}} s+\psi_P(s)\right)
= -\frac{R_2^2}{2V_{P,W}}. \Label{6-16-1}
\end{align}

Next, we set $P$ as $P_{M-}$.
Then, the Gallager bound yields 
\begin{align}
C^{\dm}_p(R_2,C^{\dm}_W|W) \le e^{-\frac{R_2^2}{2V_W^-}} \Label{Gallager2}
\end{align}
for any $R_2 <0$.
That is, 
the gap between our evaluation and the Gallager bound is equal to
the difference between 
$F(\frac{R_2}{\sqrt{V_W^-}})= \int_{-\infty}^{\frac{R_2}{\sqrt{V_W^-}}}
\frac{1}{\sqrt{2\pi}}e^{-x^2/2}dx $ and
$e^{-\frac{R_2^2}{2V_W^-}}$.
Although the former is smaller than the latter, 
both exponential rates coincide in the limit $R_2 \to \infty$.
Since we can consider that the Gallager bound gives the trivial bound for $R_2>0$, both evaluations are illustrated in Fig. \ref{graph2}.

\begin{figure}[htbp]
\begin{center}
\scalebox{1.0}{\includegraphics[scale=1]{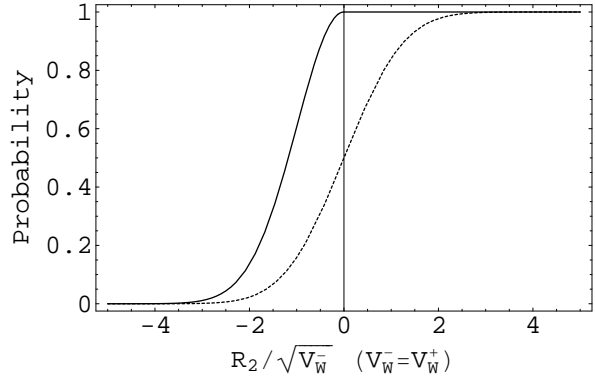}}
\end{center}
\caption{Comparison between the present evaluation and the Gallager bound.
The solid line indicates the Gallager bound, and the dotted line indicates the present evaluation.}
\Label{graph2}
\end{figure}%

Next, we consider the same comparison for the additive Markovian case.
The Gallager bound is given by
\begin{align*}
\min_{\Phi:|\Phi|\le e^{n R}}  P_{e,W(Q)^{n}}(\Phi)
\le \min_{P} \min_{0 \le s \le 1} e^{n (Rs +\psi_{Q,n}(s))},
\end{align*}
where
\begin{align*}
\psi_{Q,n} (s)\defeq &
-s \log d +\frac{1+s}{n} \log 
(
\sum_{\vec{x}_n}
Q^n(\vec{x}_n)^{\frac{1}{1+s}}).
\end{align*}
Since the asymptotic first and second cummulants of the random variable $\log Q^n(\vec{x}_n)$ 
are $-H(Q)$ and $V(Q)$, we have
\begin{align*}
\log 
(
\sum_{\vec{x}_n}
Q^n(\vec{x}_n)^{1+\frac{t}{\sqrt{n}}}) =
-H(Q) t \sqrt{n}+\frac{V(Q)}{2}t^2 + o(t^2)
\end{align*}
as $t \to 0$.
Thus,
\begin{align*}
n \psi_{Q,n} (\frac{t}{\sqrt{n}})=
(- \log d +H(Q))
t \sqrt{n} 
+ \frac{V(Q)}{2} t^2 + o(t^2).
\end{align*}
Substituting $nC_W + \sqrt{n} R_2$ and $\frac{t}{\sqrt{n}}$
into $nR$ and $s$,
we have
\begin{align*}
&C_Q^{\am}s + \frac{R_2}{\sqrt{n}} s+\psi_{Q,n}(s) \\
=& n(C_Q^{\am} \frac{t}{\sqrt{n}} + \frac{R_2}{\sqrt{n}} \frac{t}{\sqrt{n}} +\psi_{Q,n}(\frac{t}{\sqrt{n}}))\\
= &
R_2 t + \frac{V(Q)}{2} t^2 +o(t^2) \\
=& \frac{V(Q)}{2} (t+\frac{R_2}{V(Q)})^2 - \frac{R_2^2}{2 V(Q)}.
+o(t^2)
\end{align*}
Therefore, when $R_2 < 0$, 
choosing $s=\frac{-R_2}{V(Q) \sqrt{n}}$,
we obtain
\begin{align*}
\min_{\Phi:|\Phi|\le e^{n C^{\am}_Q + \sqrt{n} R_2}}  P_{e,W(Q)^{n}}(\Phi)
\le e^{-\frac{R_2^2}{2 V(Q)}},
\end{align*}
which has the same form as (\ref{Gallager2}).

In both cases, when $-3 \le R_2 \le 2$, the difference is not so small.
In such a case, it is better to use the present evaluation.
That is, the Gallager bound does not give the best evaluation in this case.
This conclusion is opposite to the exponential evaluation 
when the rate is greater than the critical rate.
Han \cite{Han1} calculated the exponential rate of the present bound, and found that it is worse than that of the Gallager bound\footnote{This description was provided in the original Japanese version, but not in the English translation.}.

Moreover, a similar conclusion was obtained in the LDPC case.
Kabashima and Saad \cite{Kaba} compared the Gallager upper bound of the average error probability and the approximation of the average error probability 
by the replica method. That is, they compared both thresholds of the rate, i.e.,
both maximum transmission rates at which the respective error probability goes to zero.
In their study (Table 1 of \cite{Kaba}), they pointed out that there exists a non-negligible difference between these two thresholds in the LDPC case. This information may be helpful for discussing the performance of the Gallager bound.

\section{Properties of $V_W^+$ and $V_W^-$}\Label{s6}

\subsection{Example}\Label{s6a}
In this section, we consider a typical example, in which,
$V_W^+$ is different than $V_W^-$.
For this purpose, 
we choose two parameters $q_1, q_2 \in [0,1]$
satisfying 
\begin{align}
&0 \le 2 q_1-q_2\le 1 \nonumber \\
&h(q_1) - \frac{h(q_2)+h(2q_1-q_2)}{2}
\le - \log \max\{q_1,1-q_1\},\Label{6-12-1}
\end{align}
where
$h(x) \defeq -x\log x -(1-x)\log (1-x)$.
According to the following three conditions (i), (ii) and (iii),
we define the 
five joint distributions 
$W_1$, $W_2$, $W_3$, $W_4$, and $W_5$ 
on two random variables $A=0,1$ and $B=0,1$.
In the following, $Q^A$ ($Q^B$) denotes the marginal distribution of $A$ concerning $A$ ($B$).
\begin{description}
\item[(i) Uniformity on $A$] \quad\par
All distributions are assumed to satisfy
\begin{align*}
W_i^A(0)=1/2.
\end{align*}
\item[(ii) Same marginal distribution on $B$ for $i=1,2$] \quad\par
Two random variables $A=0,1$ and $B=0,1$ are not independent
in $W_1$ and $W_2$, but 
$W_1$ and $W_2$ have the same marginal distribution on $B$.
That is,
\begin{align*}
W_1^B(0|A=0)&=W_2^B(0|A=1)=q_2\\
W_1^B(0|A=1)&=W_2^B(0|A=0)=2q_1-q_2.
\end{align*}
Thus, $W_1$ and $W_2$ satisfy
\begin{align*}
W_1^B(0)=W_2^B(0)=q_1.
\end{align*}
\item[(iii) Independence between $A$ and $B$ for $i=3,4,5$] \quad\par
Due to the condition (\ref{6-12-1}),
there exist two solutions for $x$ in the following equation
because $d(x\|q_1)$ is monotone increasing in $(q_1,1)$ and 
is monotone decreasing in $(0,q_1)$:
\begin{align*}
h(q_1)- \frac{h(q_2)+h(2q_1-q_2)}{2}
= d(x\|q_1),
\end{align*}
where 
\begin{align*}
d(x\|y) &\defeq x \log \frac{x}{y}+
(1-x) \log \frac{1-x}{1-y}.
\end{align*}
Letting $p_1$ and $p_2$ be these two solutions,
we define three distributions
$W_3$, $W_4$, and $W_5$, in which 
two random variables $A=0,1$ and $B=0,1$ are independent,
by 
\begin{align*}
W_3^B(0)=p_1,~
W_4^B(0)=p_2,~
W_5^B(0)=q_1.
\end{align*}
\end{description}
From the construction, we can check that 
\begin{align}
D(W_i\|W_5)=h(q_1)- \frac{h(q_2)+h(2q_1-q_2)}{2}\Label{12-13-1}
\end{align}
for $i=1,2,3,4$.
Consider the subsets
\begin{align*}
{\cal Z}_0 &\defeq
\{
Q|
Q^A(0)=1/2\}\\
{\cal Z}_1 &\defeq
\{
Q\in {\cal Z}_0|
Q^B(0)=q_1\}\\
{\cal Z}_2 &\defeq
\{
Q\in {\cal Z}_0|
Q^B(0|A=0)=Q^B(0|A=1)\}.
\end{align*}
Then, ${\cal Z}_1 \cap {\cal Z}_2=\{W_5\}$.
Hence, the relationship among ${\cal Z}_0$, ${\cal Z}_1$, ${\cal Z}_2$, 
$W_1$, $W_2$, $W_3$, $W_4$, and $W_5$ is shown in Fig. \ref{pic1}.
For any distribution $Q$,

\begin{figure}[htbp]
\begin{center}
\scalebox{1.0}{\includegraphics[scale=0.5]{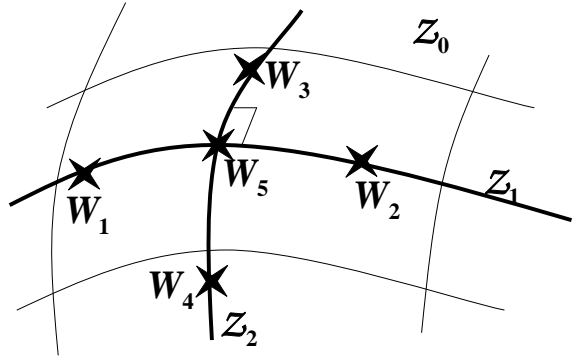}}
\end{center}
\caption{${\cal Z}_0$, ${\cal Z}_1$, ${\cal Z}_2$, 
$W_1$, $W_2$, $W_3$, $W_4$, and $W_5$}
\Label{pic1}
\end{figure}%

Then, the following lemma holds.
\begin{lem}\Label{lem5}
\begin{align}
\argmax_{Q}\min_{x=1,2}D(W_x\|Q)&= 
\argmax_{Q \in {\cal Z}_1}\min_{x=1,2}D(W_x\|Q)\Label{6-13-1}\\
\argmax_{Q}\min_{x=3,4}D(W_x\|Q)&= 
\argmax_{Q \in {\cal Z}_2}\min_{x=3,4}D(W_x\|Q). \Label{6-13-2}
\end{align}
\end{lem}

Therefore, (\ref{12-13-1}) implies that
\begin{align*}
\argmax_{Q}\min_{x=1,2,3,4}D(W_x\|Q)=W_5.
\end{align*}
and
\begin{align*}
\max_{Q}\min_{x=1,2}D(W_x\|Q)&= 
\max_{Q \in {\cal Z}_1}\min_{x=1,2}D(W_x\|Q)=h(q_1)- \frac{h(q_2)+h(2q_1-q_2)}{2}\\
\max_{Q}\min_{x=3,4}D(W_x\|Q)&= 
\max_{Q \in {\cal Z}_2}\min_{x=3,4}D(W_x\|Q)=h(q_1)- \frac{h(q_2)+h(2q_1-q_2)}{2}.
\end{align*}
That is, the capacity of the channel $x=1,2,3,4 \mapsto W_x$ is calculated as
\begin{align*}
C^{\dm}_W=\max_{Q}\min_{x=1,2,3,4}D(W_x\|Q)= h(q_1)- \frac{h(q_2)+h(2q_1-q_2)}{2}.
\end{align*}
Then, the set ${\cal V}$ is given by the convex hull of 
$P=(1/2,1/2,0,0)$ and $P'=(0,0,\frac{q_1-p_2}{p_1-p_2},\frac{q_1-p_1}{p_2-p_1})$.
Thus,
$V_{\lambda P+(1-\lambda)P',W}
=\lambda V_{P,W}
+(1-\lambda)V_{P',W}$.
When 
$V_{P,W} \le V_{P',W}$,
\begin{align*}
V_W^+=V_{P',W},
V_W^-=V_{P,W}.
\end{align*}
Otherwise,
\begin{align*}
V_W^+=V_{P,W},
V_W^-=V_{P',W}.
\end{align*}
Our numerical analysis (Fig. \ref{graph1})
suggests the relation $V_{P,W} \le V_{P',W}$.

\begin{figure}[htbp]
\begin{center}
\scalebox{1.0}{\includegraphics[scale=1]{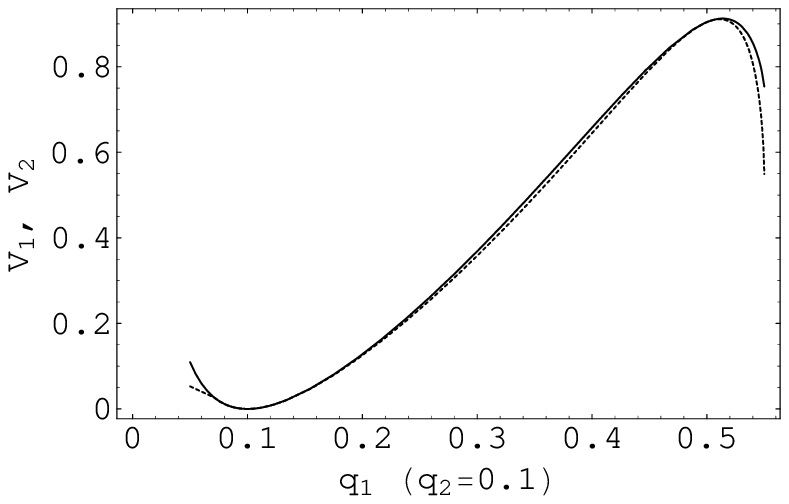}}
\scalebox{1.0}{\includegraphics[scale=1]{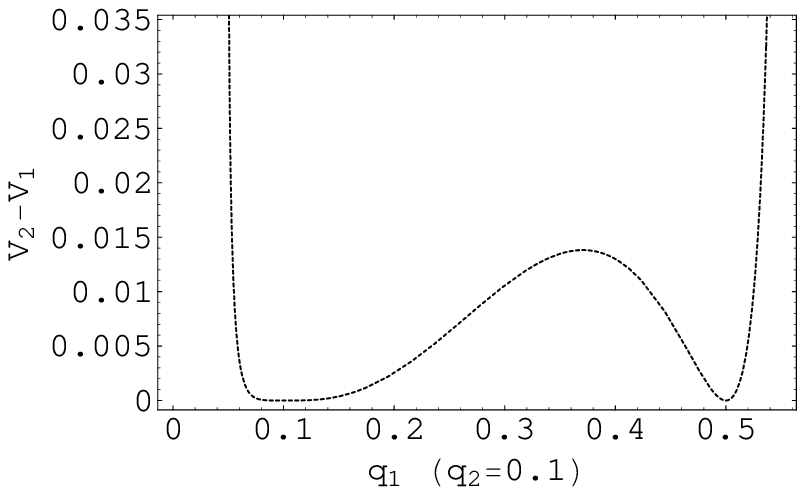}}
\end{center}
\caption{Comparison between $V_1=V_{P,W}$ (dotted line) and $V_2=V_{P',W}$ (solid line).}
\Label{graph1}
\end{figure}%

\noindent{\it Proof of Lemma \ref{lem5}:\quad}
For this proof, we define the maps ${\cal E}_A$ and ${\cal E}_B$ as
\begin{align*}
({\cal E}_A Q)(A=a,B=b):= & P^A(a)  Q(B=b|A=a) \\
({\cal E}_B Q)(A=a,B=b):= & P^B(b)  Q(A=a|B=b) ,
\end{align*}
where $P^A(0)=1/2$ and $P^B(0)=q_1$.
when the distribution $Q'$ satisfies that ${Q'}^A=P^A$,
the following Pythagorean type inequality
\begin{align}
D(Q'\|Q)= D(Q'\|{\cal E}_A (Q))+  D({\cal E}_A (Q)\| Q)
\Label{6-13-4}
\end{align}
holds.
Similarly, when the distribution $Q'$ satisfies that ${Q'}^B=P^B$,
the following Pythagorean type inequality
\begin{align}
D(Q'\|Q)= 
D(Q'\|{\cal E}_B (Q))+  D({\cal E}_B Q \|Q)\Label{6-13-4}
\end{align}
holds.
Define 
$Q_{2k}:= \underbrace{{\cal E}_B \circ{\cal E}_A \circ \cdots \circ {\cal E}_B \circ{\cal E}_A}_{2k}Q$
and 
$Q_{2k+1}:= \underbrace{{\cal E}_A \circ {\cal E}_B \circ{\cal E}_A \circ \cdots \circ {\cal E}_B \circ{\cal E}_A}_{2k+1}Q$.
Then, 
$D(Q_{2k+1}\|Q_{2k}) =
D({\cal E}_A Q_{2k}\|{\cal E}_A Q_{2k-1}) \le
D(Q_{2k}\|Q_{2k-1}) $, and 
$D(Q_{2k}\|Q_{2k-1}) \le
D(Q_{2k-1}\|Q_{2k-2}) $.
For any $Q' \in {\cal Z}_1$, we have
\begin{align*}
D(Q'\|Q)= D(Q'\|Q_n) + \sum_{k=1}^n D(Q_{k}\|Q_{k-1}) .
\end{align*}
Thus, 
$D(Q_{k}\|Q_{k-1})$ converges to zero.
Therefore, there exists a distribution $Q_{\infty}$ such that
$Q_k \to Q_{\infty}$.
Hence,
\begin{align*}
D(Q'\|Q)= D(Q'\|Q_\infty) + \sum_{k=1}^\infty D(Q_{k}\|Q_{k-1}) ,
\end{align*}
which implies (\ref{6-13-1}).

Further, for any $P_2 \in {\cal Z}_2$, 
we assume that $Q$ satisfies $Q^A=P^A$.
Since
the concavity of $\log$ implies the inequality
$\log \sum_a P^A(a) Q(B=b|A=a)
\ge \sum_a P^A(a) \log  Q(B=b|A=a)$,
the following Pythagorean type inequality
\begin{align}
& D(P_2\|Q)= 
H(P_2)-\sum_a \sum_b P_2^A(a) P_2^B(b) \log Q(a,b) \nonumber \\
= &
H(P_2)
-\sum_a P_2^A(a) \log Q^A(a)
-\sum_a \sum_b P_2^A(a) P_2^B(b) \log Q(B=b|A=a) \nonumber\\
= &
H(P_2)
-\sum_a P_2^A(a) \log Q^A(a)
-\sum_b P_2^B(b) \log Q^B(b)
+\sum_b P_2^B(b) \log Q^B(b)
-\sum_a \sum_b P_2^A(a) P_2^B(b) \log Q(B=b|A=a)\nonumber\\
= &
D(P_2\|P_2^A \times P_2^B)
+\sum_b P_2^B(b) \log Q^B(b)
-\sum_b P_2^B(b)
\sum_a P_2^A(a) \log Q(B=b|A=a)\nonumber\\
= &
D(P_2\|P_2^A \times P_2^B)
+\sum_b P_2^B(b) 
\left(\log \sum_a P^A(a) Q(B=b|A=a)
- \sum_a P^A(a) \log  Q(B=b|A=a)\right) \nonumber\\
\ge &
D(P_2\|P_2^A \times P_2^B)\Label{6-13-3}
\end{align}
holds.
Combination of (\ref{6-13-4}) and (\ref{6-13-3}) yields (\ref{6-13-2}).

\endproof

\subsection{Additivity}\Label{s6b}
The capacity satisfies the additivity condition.
That is, for any two channels $\{W_x(y)\}$ and $\{W_{x'}'(y')\}$,
the combined channel $\{(W\times W')_{x,x'}(y,y')=W_x(y)W_{x'}'(y')\}$ 
satisfies the following:
\begin{align*}
C^{\dm}_{W\times W'}=C^{\dm}_W+C^{\dm}_{W'}.
\end{align*}
Similarly,
as mentioned in the following lemma,
$V_W^+$ and $V_W^-$ satisfy the additivity condition.
\begin{lem}\Label{l3}
The equations
\begin{align}
V_{W\times W'}^+&=
V_W^++V_{W'}^+\Label{12-11-7}\\
V_{W\times W'}^-&=
V_W^-+V_{W'}^-\Label{12-11-8}
\end{align}
hold.
\end{lem}

\noindent{\it Proof of Lemma \ref{l3}:\quad}
We choose the distributions $Q$ and $Q'$ as
\begin{align*}
Q &\defeq \argmin_Q \max_x D(W_x\| Q) \\
Q' &\defeq \argmin_{Q'} \max_{x'} D(W_{x'}'\| Q') .
\end{align*}
Then, 
\begin{align*}
Q \times Q'= \argmin_{Q''} \max_{x,x'} D(W_x\times W'_{x'}\| Q'').
\end{align*}
Assume that a distribution $P$ with the random variables $x$ and $x'$
satisfies the following:
\begin{align}
\sum_{x,x'}P(x,x')W_x \times W_{x'}'&=Q\times  Q',\Label{12-11-6}\\
I(P,W\times W')&=C^{\dm}_W+C^{\dm}_{W'}.\Label{12-13-5}
\end{align}
Then,
the marginal distributions $P_1$ and $P_1$ of $P$ concerning $x$ and $x'$
satisfy
\begin{align*}
I(P_1,W)=C^{\dm}_W ,~
I(P_2,W')=C^{\dm}_{W'},
\end{align*}
which implies 
\begin{align*}
D(W_x\|Q)=C^{\dm}_W,~
D(W'_{x'}\|Q')=C^{\dm}_{W'}
\end{align*}
for $x \in \supp(P_1)$ and $x' \in \supp(P_2)$,
where $\supp(P)$ denotes the support of the distribution $P$.
Hence,

\begin{align*}
& V_{P,W\times W'}=
\sum_{x,x'}P(x,x')\sum_{y,y'}W_x(y)W_{x'}'(y')
(\log \frac{W_x(y)}{Q(y)}+ \log \frac{W_{x'}'(y')}{Q'(y')})^2
-(D(W_x\|Q)+D(W_{x'}'\|Q'))^2\\
=&
\sum_{x,x'}P(x,x')\sum_{y,y'}W_x(y)W_{x'}'(y')
\left(
(\log \frac{W_x(y)}{Q(y)})^2+ (\log \frac{W_{x'}'(y')}{Q'(y')})^2
+2  \log \frac{W_x(y)}{Q(y)}\log \frac{W_{x'}'(y')}{Q'(y')}
\right)\\
&-(
D(W_x\|Q)^2+D(W_{x'}'\|Q')^2+ 2D(W_x\|Q) D(W_{x'}'\|Q')
)\\
=&
\sum_{x,x'}P(x,x')\sum_{y,y'}W_x(y)W_{x'}'(y')
\left(
(\log \frac{W_x(y)}{Q(y)})^2+ (\log \frac{W_{x'}'(y')}{Q'(y')})^2
\right)
-D(W_x\|Q)^2
-D(W_{x'}'\|Q')^2
\\
=&V_{P_1,W}+V_{P_2,W'}.
\end{align*}

Therefore, when the conditions (\ref{12-11-6}) and (\ref{12-13-5}) are satisfied, the maximum of $V_{P,W\times W'}$ is equal to 
$V_W^++V_{W'}^+$, which implies (\ref{12-11-7}).
Similarly, we obtain (\ref{12-11-8}).
\endproof

The same fact holds with the cost constraint.
The capacity with the cost constraint satisfies the additivity condition.
That is, for any two cost fucntions $c$ and $c'$ for channels $\{W_x(y)\}$ and $\{W_{x'}'(y')\}$,
the combined cost $(c+c')(x,x')\defeq c(x)+c'(x')$ 
satisfies the following:
\begin{align*}
C^{\dm}_{W\times W',c+c',K+K'}=C^{\dm}_{W,c,K}+C^{\dm}_{W',c',K'}.
\end{align*}
The quantities $V_{W,c,K}^+$ and $V_{W,c,K}^-$ satisfy the additivity condition.
\begin{lem}\Label{l3b}
The equations
\begin{align}
V_{W\times W',c+c',K+K'}^+&=
V_{W,c,K}^++V_{W',c',K'}^+\Label{12-11-7b}\\
V_{W\times W',c+c',K+K'}^-&=
V_{W,c,K}^-+V_{W',c',K'}^-\Label{12-11-8b}
\end{align}
hold.
\end{lem}
This lemma can be proven in the same way as Lemma \ref{l3} by replacing the definitions of $Q$ and $Q'$ by 
\begin{align*}
Q &\defeq \argmin_Q \max_{P:{\rm E}_Pc(x)\le K} \sum_x P(x) D(W_x\| Q) \\
Q' &\defeq \argmin_{Q'} \max_{P':{\rm E}_{P'}c'(x')\le K'} \sum_{x'} P'(x') D(W_{x'}'\| Q') .
\end{align*}

\section{Notations of the information spectrum}\Label{s7}
\subsection{Information Spectrum}
In the present paper, we treat general channels.
First, we focus on two sequences of 
probability spaces $\{{\cal X}_n\}_{n=1}^\infty$ of the input signal and
those $\{{\cal Y}_n\}_{n=1}^\infty$ of the output signal,
and a sequence of probability 
transition matrixes $\specW\defeq
\{W^n(y|x)\}_{n=1}^\infty$.
We also focus on 
a sequence of distributions on input systems
$\specP\defeq\{P^n\}_{n=1}^{\infty}$.
The asymptotic behavior of the logarithmic likelihood ratio between
$W^n_x(y)\defeq W^n(y|x)$ and $W^n_{P^n}(y)\defeq\sum_{x\in {\cal X}_n}P^n(x)W^n(y|x)$
can be characterized by the following quantities 
\begin{align*}
{I}_p(R|\specP,\specW)
&\defeq 
\limsup_{n \to \infty}
\sum_{x \in {\cal X}_n} P^n(x) W^n_x
\left\{
\frac{1}{n}\log \frac{W^n_x(y)}{W^n_{P^n}(y)}
< R
\right\}\\
{I}(\epsilon|\specP,\specW)
&\defeq 
\sup
\{ R |
{I}_p(R|\specP,\specW)\le \epsilon
\}\\
&= 
\inf
\{ R |
{I}_p(R|\specP,\specW)\ge \epsilon
\}
\end{align*}
for $0 \le \epsilon < 1$.
Focusing on a sequence of distributions on output systems
$\specQ\defeq\{Q^n\}_{n=1}^{\infty}$,
we can define 
\begin{align*}
{J}_p(R|\specP,\specQ,\specW)
&\defeq 
\limsup_{n \to \infty}
\sum_{x \in {\cal X}_n} P^n(x) W^n_x
\left\{
\frac{1}{n}\log \frac{W^n_x(y)}{Q^n(y)}
< R
\right\}\\
{J}(\epsilon|\specP,\specQ,\specW)
&\defeq 
\sup
\{ R |
{J}_p(R|\specP,\specQ,\specW)\le \epsilon
\}\\
&= 
\inf
\{ R |
{J}_p(R|\specP,\specQ,\specW)\ge \epsilon
\}
\end{align*}
for $0 \le \epsilon < 1$.

When the channel $W^n$ is the $n$-th stationary discrete memoryless channel  $W^{\times n}$ of $W(y|x)$ and the probability distribution $\specP=\{P^n\}$ is the $n$-th independent and identical distribution $P^{\times n}$ of $P$, the law of large numbers guarantees that ${I}(\epsilon|\specP,\specW)$ coincides with the mutual information $I(P,W) = \sum_{x,y} P(x) W_x(y) \log \frac{W_x(y)}{W_P(y)}$.
For a more detailed description of asymptotic behavior, 
we focus on the second order of the coding length $n^{\beta}$ for $\beta <1$.
In order to characterize the coefficient of the second order $n^{\beta}$,
we introduce the following quantities:
\begin{align*}
{I}_p(R_2,R_1|\specP,\specW)
&\defeq 
\limsup_{n \to \infty}
\sum_{x \in {\cal X}_n} P^n(x) W^n_x
\left\{
\frac{1}{n^{\beta}}
(\log \frac{W^n_x(y)}{W^n_{P^n}(y)}-n R_1)
< R_2
\right\}\\
{I}(\epsilon,R_1|\specP,\specW)
&\defeq 
\sup
\{ R_2 |
{I}_p(R_2,R_1|\specP,\specW)\le \epsilon
\}\\
&= 
\inf
\{ R_2 |
{I}_p(R_2,R_1|\specP,\specW)\ge \epsilon
\}
\end{align*}
for $0 \le \epsilon <1$.
Similarly, 
${J}_p(R_2,R_1|\specP,\specQ,\specW)$ and
${J}(\epsilon,R_1|\specP,\specQ,\specW)$
are defined for
$0 \le \epsilon < 1$.
When $\specW$ is $\specWt=\{W^{\times n}\}$ and
$\specP$ is $\specPt=\{P^{\times n}\}$,
the second order of the coding length is $n^{\frac{1}{2}}$
and
the central limit theorem guarantees that
$\frac{1}{n^{\frac{1}{2}}}
(\log \frac{W^n_x(y)}{W^n_{P^n}(y)}-n 
I(P,W))$
asymptotically obeys the Gaussian distribution with expectation $0$ and variance:
\begin{align*}
V_{P,W} \defeq \sum_{x}P(x) \sum_y W_x(y)
\left(\log \frac{W_x(y)}{W_{P}(y)}-I(P,W)\right)^2.
\end{align*}
Therefore, using the distribution function $F$ for the standard Gaussian distribution,
we can express the above quantities as follows:
\begin{align}
{I}(\epsilon,I(P,W)|\specPt,\specWt)
= \sqrt{V_{P,W}}\erf^{-1}(\epsilon) \Label{25}.
\end{align}

In the case of additive channels, we focus on the limiting behavior of 
the entropy rate of the distributions $\specQ=\{Q^n\}_{n=1}^{\infty}$ describing the additive noise.
Similar to the above, we define the following.
\begin{align*}
{H}_p(R|\specQ)
&\defeq 
\liminf_{n \to \infty}
\sum_{x \in {\cal X}_n} Q^n
\left\{\frac{-1}{n}\log Q^n(x)< R\right\}\\
{H}(\epsilon|\specQ)
&\defeq 
\sup
\{ R |
{H}_p(R|\specQ)\le \epsilon
\}\\
&= 
\inf
\{ R |
{H}_p(R|\specQ)\ge \epsilon
\}\\
{H}_p(R_2,R_1|\specQ)
&\defeq 
\liminf_{n \to \infty}
\sum_{x \in {\cal X}_n} Q^n
\left\{\frac{1}{n^{\beta}}
(-\log Q^n(x)-n R_1)
< R_2
\right\}\\
{H}(\epsilon,R_1|\specQ)
&\defeq 
\sup
\{ R_2 |
{H}_p(R_2,R_1|\specQ)\le \epsilon
\}\\
&= 
\inf
\{ R_2 |
{H}_p(R_2,R_1|\specQ)\ge \epsilon
\}
\end{align*}
for $0 \le \epsilon <1$.
As is discussed in Section VII in \cite{H-sec}, when $\specQ$ is given by a Markovian process $Q(y|x)$,
the relationships
\begin{align}
{H}(\epsilon|\specQ)&= H(Q) \Label{6-11-11}\\
{H}(\epsilon,H(Q)|\specQ) &= \sqrt{V(Q)}G^{-1}(\epsilon) \Label{6-11-12} \\ 
{H}_p(R_2,H(Q)|\specQ)&
=G(R_2/\sqrt{V(Q)}) \Label{6-11-13}
\end{align}
hold with $\beta=1/2$.

\subsection{Stochastic limits}
In order to treat the relationship between the above quantities, we consider the limit superior in probability $\plimsup_{n \to \infty}$ and the limit inferior in probability $\pliminf_{n \to \infty}$, which are defined by
\begin{align*}
\plimsup_{n \to \infty} Z_n|_{P_n} &\defeq \inf\{ a | \lim_{n \to \infty} P_n\{Z_n>a\}=0\}\\
\pliminf_{n \to \infty} Z_n|_{P_n} &\defeq \sup\{ a | \lim_{n \to \infty} P_n\{Z_n<a\}=0\}.
\end{align*}
In particular, when $\plimsup_{n \to \infty} Z_n|_{P_n} = \pliminf_{n \to \infty} Z_n|_{P_n}=a$,
we write
\begin{align*}
\plim_{n \to \infty} Z_n|_{P_n} =a.
\end{align*}
The concept $\pliminf_{n \to \infty}$ can be generalized as
\begin{align*}
\epliminf_{n \to \infty} Z_n|_{P_n}\defeq \sup\{ a | \limsup_{n \to \infty} P_n\{Z_n<a\}
\le \epsilon \}.
\end{align*}
From the definitions, we can check the following properties:
\begin{align}
\epliminf_{n \to \infty} Z_n+Y_n |_{P_n}
\ge &
\epliminf_{n \to \infty} Z_n |_{P_n}
+ \pliminf_{n \to \infty} Y_n |_{P_n}.\Label{12-4-1}\\
\epliminf_{n \to \infty} Z_n+Y_n |_{P_n}
\le &
\epliminf_{n \to \infty} Z_n |_{P_n}
+ \plimsup_{n \to \infty} Y_n |_{P_n}.\Label{6-11-8}
\end{align}
As shown by Han \cite{Han1}, the relation
\begin{align}
\pliminf_{n \to \infty} \frac{1}{n^{\alpha}}
\left.\log\frac{P^n(x)}{{P^n}'(x)}
\right|_{P^n} 
& \ge 0\Label{12-4-2}
\end{align}
holds for $\alpha >0$ and any two sequences $\specP=\{P^n\}$ 
and $\specPd=\{{P^n}'\}$ of distributions with the variable $x$.

By using this concept, 
${I}(\epsilon|\specP,\specW)$,
${J}(\epsilon|\specP,\specQ,\specW)$,
${I}(\epsilon,R_1|\specP,\specW)$, and
${J}(\epsilon,R_1|\specP,\specQ,\specW)$
are characterized by
\begin{align*}
{I}(\epsilon|\specP,\specW)
&=\left. \epliminf_{n \to \infty} 
\frac{1}{n}\log \frac{W^n_x(y)}{W^n_{P^n}(y)}\right|_{\rP_{P^n,W^n}}\\
{J}(\epsilon|\specP,\specQ,\specW)
&=\left.\epliminf_{n \to \infty} 
\frac{1}{n}\log \frac{W^n_x(y)}{Q^n(y)}\right|_{\rP_{P^n,W^n}}\\
{I}(\epsilon,R_1|\specP,\specW)
&=\left.\epliminf_{n \to \infty} 
\frac{1}{n^{\beta}}(\log \frac{W^n_x(y)}{W^n_{P^n}(y)}
-nR_1)\right|_{\rP_{P^n,W^n}}\\
{J}(\epsilon,R_1|\specP,\specQ,\specW)
&=\left.\epliminf_{n \to \infty} 
\frac{1}{n^{\beta}}(\log \frac{W^n_x(y)}{Q^n(y)}
-nR_1)\right|_{\rP_{P^n,W^n}}.
\end{align*}
Substituting $W^n_{P^n}$ and $Q^n$ into $P^n$ and ${P^n}'$ in (\ref{12-4-2}),
and using (\ref{12-4-1}),
we obtain
\begin{align*}
{I}(\epsilon|\specP,\specW)
&\le {J}(\epsilon|\specP,\specQ,\specW)\\
{I}(\epsilon,R_1|\specP,\specW)
&\le
{J}(\epsilon,R_1|\specP,\specQ,\specW).
\end{align*}
Since $1-H_p(R|\specQ)=\liminf_{n \to \infty} Q^n\{
\frac{1}{n}\log Q^n(x) < -R\}$, 
$H(\epsilon|\specQ)$ is characterized as
\begin{align}
& -H(\epsilon|\specQ)
= - \inf \{R| H_p(R|\specQ) \ge \epsilon \}\nonumber \\
= & \sup \{-R| 1- H_p(R|\specQ) \le 1-\epsilon \}
= \eepliminf_{n \to \infty}  \left.\frac{1}{n}\log Q^n(x)\right.|_{Q^n}.\Label{6-11-9}
\end{align}
Similarly, 
\begin{align}
-H(\epsilon,R_1|\specQ)
= \eepliminf_{n \to \infty} \left.\frac{1}{n^{\beta}}(\log Q^n(x)+n R_1)\right|_{Q^n}.\Label{6-11-10b}
\end{align}
In the following, we discuss 
the relationship between the above-mentioned quantities and channel capacities. 


\section{General asymptotic formulas}\Label{s8}
\subsection{General case}
Next, we consider the $\epsilon$ capacity and its related quantity, which are defined by
\begin{align*}
C_p(R|\specW) &\defeq 
\inf_{\{\Phi_n\}_{n=1}^{\infty}
}
\left\{
\left. \limsup_{n \to \infty} P_{e,W^n}(\Phi_n )
\right|
\liminf_{n \to \infty}
\frac{1}{n} \log |\Phi_n| \ge R
\right\} \\
C(\epsilon|\specW) &\defeq 
\sup_{\{\Phi_n\}_{n=1}^{\infty}
}
\left\{
\left.\liminf_{n \to \infty}
\frac{1}{n}\log |\Phi_n| 
\right|
\limsup_{n \to \infty} P_{e,W^n}(\Phi_n ) \le \epsilon
\right\}.
\end{align*}
Concerning these quantities, 
the following general asymptotic formulas hold.
\begin{thm}\Label{thm3}(Verd\'{u} \& Han\cite{Verdu-Han}, Hayashi \& Nagaoka \cite{H-N})
The relations
\begin{align}
C_p(R|\specW)&=
\inf_{\specP}
\lim_{\gamma \downarrow 0}{I}_p(R-\gamma|\specP,\specW) 
=\inf_{\specP}\sup_{\specQ} \lim_{\gamma \downarrow 0}
{J}_p(R-\gamma|\specP,\specQ,\specW)
\Label{11-30-1}\\
C(\epsilon|\specW)&=
\sup_{\specP}I(\epsilon|\specP,\specW)
=\sup_{\specP}\inf_{\specQ} J(\epsilon|\specP,\specQ,\specW)
\Label{11-30-2}
\end{align}
hold for $0\le \epsilon < 1$.
\end{thm}
\begin{rem}\rm
Historically,
Verd\'{u} \& Han \cite{Verdu-Han} proved the first equation in (\ref{11-30-2}).
Hayashi \& Nagaoka \cite{H-N} established the second equation in (\ref{11-30-2}) with $\epsilon =0$ for the first time, even for the classical case, 
although their main topic was the quantum case.
The relation (\ref{11-30-1}) is proven for the first time in this paper.
\end{rem}

Next, we proceed to the second-order coding rate.
As a generalization of (\ref{11-30-3}) and (\ref{11-30-4}),
we define the following:
\begin{align}
C_p(R_2,R_1|\specW)& \defeq 
\inf_{\{\Phi_n\}_{n=1}^{\infty}
}
\left\{\left.
\limsup_{n \to \infty} P_{e,W^n}(\Phi_n )
\right|
\liminf_{n \to \infty}
\frac{1}{n^{\beta}}(\log |\Phi_n|- nR_1) \ge R_2
\right\} \Label{11-30-5} \\
C(\epsilon, R_1|\specW)& \defeq 
\sup_{\{\Phi_n\}_{n=1}^{\infty}
}
\left\{\left.
\liminf_{n \to \infty}
\frac{1}{n^{\beta}}(\log |\Phi_n|- nR_1) 
\right|
\limsup_{n \to \infty} P_{e,W^n}(\Phi_n ) \le \epsilon
\right\}.\Label{11-30-6}
\end{align}
Similar to Theorem \ref{thm3}, the following general formulas for the second-order coding rate hold.
\begin{thm}\Label{thm4}
The relations
\begin{align}
C_p(R_2,R_1|\specW)&=
\inf_{\specP}\lim_{\gamma \downarrow 0} {I}_p(R_2-\gamma,R_1|\specP,\specW)
=\inf_{\specP}\sup_{\specQ} \lim_{\gamma \downarrow 0} {J}_p(R_2-\gamma,R_1|\specP,\specQ,\specW)
\Label{11-30-7}\\
C(\epsilon,R_1|\specW)&=
\sup_{\specP}I(\epsilon,R_1|\specP,\specW)
=\sup_{\specP}\inf_{\specQ} J(\epsilon,R_1|\specP,\specQ,\specW)
\Label{11-30-8}
\end{align}
hold for $0\le \epsilon < 1$.
\end{thm}

Indeed, Theorem \ref{thm4} has greater significance than generalization.
This theorem provides a unified viewpoint concerning the second order asymptotic rate in channel coding
and the following merits.
First, it shortens the proof of 
Theorem \ref{thm2b}.
Second it enables us to extend Theorem \ref{thm2b} to the case of cost constraint.
Third, it yields the extension to Gaussian noise case, which has continuous input signals.
Fourth, it allows us to extend the same treatment to the Markovian case with the additive noise.

\subsection{Cost constraint}
We focus on a sequence of cost function $\specc=\{c_n\}_{n=1}^\infty$ where $c_n$
is a function from ${\cal X}_n$ to $\real$.
In this case, all alphabets are assumed to belong to the set 
\begin{align*}
{\cal X}_{n,c,K}\defeq 
\left\{x\in {\cal X}_n
\left|
\sum_{i=1}^n c_n(x) \le n K \right. \right\}.
\end{align*}
That is,
our code $\{\Phi_n\}$ is assumed to satisfy that
$\supp(\Phi_n)\subset {\cal X}_{n,c,K}$.
Then, the capacities with cost constraint are given by
\begin{align}
C_p(R|\specW,\specc,K) &\defeq 
\inf_{\{\Phi_n\}_{n=1}^{\infty}
}
\left\{
\left. \limsup_{n \to \infty} P_{e,W^n}(\Phi_n )
\right|
\liminf_{n \to \infty}
\frac{1}{n} \log |\Phi_n| \ge R,
\supp(\Phi_n) \subset {\cal X}_{n,c,K}
\right\} \nonumber \\
C(\epsilon|\specW,\specc,K) &\defeq 
\sup_{\{\Phi_n\}_{n=1}^{\infty}
}
\left\{
\left.\liminf_{n \to \infty}
\frac{1}{n}\log |\Phi_n| 
\right|
\limsup_{n \to \infty} P_{e,W^n}(\Phi_n ) \le \epsilon,
\supp(\Phi_n) \subset {\cal X}_{n,c,K}
\right\}\nonumber \\
C_p(R_2,R_1|\specW,\specc,K)& \defeq 
\inf_{\{\Phi_n\}_{n=1}^{\infty}
}
\left\{\left.
\limsup_{n \to \infty} P_{e,W^n}(\Phi_n )
\right|
\liminf_{n \to \infty}
\frac{1}{n^{\beta}}(\log |\Phi_n|- nR_1) \ge R_2,
\supp(\Phi_n) \subset {\cal X}_{n,c,K}
\right\}. \Label{11-30-5b} \\
C(\epsilon, R_1|\specW,\specc,K)& \defeq 
\sup_{\{\Phi_n\}_{n=1}^{\infty}
}
\left\{\left.
\liminf_{n \to \infty}
\frac{1}{n^{\beta}}(\log |\Phi_n|- nR_1) 
\right|
\limsup_{n \to \infty} P_{e,W^n}(\Phi_n ) \le \epsilon,
\supp(\Phi_n) \subset {\cal X}_{n,c,K}
\right\}.\Label{11-30-6b}
\end{align}

Concerning these quantities, 
the following general asymptotic formulas hold.
\begin{thm}\Label{thm3b}(Han\cite{Han1}, Hayashi \& Nagaoka \cite{H-N})
The relations
\begin{align}
C_p(R|\specW,\specc,K)&=
\inf_{\specP
:\supp(P_n)\subset {\cal X}_{n,c,K}
 }
\lim_{\gamma \downarrow 0}{I}_p(R-\gamma|\specP,\specW) 
=\inf_{\specP}\sup_{\specQ} \lim_{\gamma \downarrow 0}
{J}_p(R-\gamma|\specP,\specQ,\specW)
\Label{11-30-1b}\\
C(\epsilon|\specW,\specc,K)&=
\sup_{\specP:\supp(P_n)\subset {\cal X}_{n,c,K}}I(\epsilon|\specP,\specW)
=\sup_{\specP:\supp(P_n)\subset {\cal X}_{n,c,K}} \inf_{\specQ} J(\epsilon|\specP,\specQ,\specW)
\Label{11-30-2b}
\end{align}
hold for $0\le \epsilon < 1$.
\end{thm}

\begin{rem}\rm
Historically,
Han \cite{Han1} proved the first equation in (\ref{11-30-2b}).
Hayashi \& Nagaoka \cite{H-N} established the second equation in (\ref{11-30-2b}) with $\epsilon =0$ for the first time, even for the classical case, 
although their main topic was the quantum case.
The relation (\ref{11-30-1b}) is proven for the first time in this paper.
\end{rem}
Similar to Theorem \ref{thm4}, the following general formulas for the second-order coding rate hold.
\begin{thm}\Label{thm4b}
The relations
\begin{align}
C_p(R_2,R_1|\specW,\specc,K)&=
\inf_{\specP:\supp(P_n)\subset {\cal X}_{n,c,K}}\lim_{\gamma \downarrow 0} {I}_p(R_2-\gamma,R_1|\specP,\specW)
=\inf_{\specP:\supp(P_n)\subset {\cal X}_{n,c,K}}\sup_{\specQ} \lim_{\gamma \downarrow 0} {J}_p(R_2-\gamma,R_1|\specP,\specQ,\specW)
\Label{11-30-7b}\\
C(\epsilon,R_1|\specW,\specc,K)&=
\sup_{\specP:\supp(P_n)\subset {\cal X}_{n,c,K}}I(\epsilon,R_1|\specP,\specW)
=\sup_{\specP:\supp(P_n)\subset {\cal X}_{n,c,K}}\inf_{\specQ} J(\epsilon,R_1|\specP,\specQ,\specW)
\Label{11-30-8b}
\end{align}
hold for $0\le \epsilon < 1$.
\end{thm}

The above theorems can be regarded as special cases of Theorems \ref{thm3} and \ref{thm4} 
by substituting the set ${\cal X}_{n,c,K}$ into the set ${\cal X}_{n}$.
Hence, it is sufficient to show Theorems \ref{thm3} and \ref{thm4}.

\subsection{Additive case}
Next, we consider the case where the channel is given as a sequence of additive channel $\specW(\specQ)=\{W^n(Q^n)(y|x)= Q^n(y-x)\}$ on the set ${\cal X}^n$ with the cardinality $d$.
Verd\'{u} \& Han proved the following theorem.
\begin{thm}\Label{thm31}(Verd\'{u} \& Han \cite{Verdu-Han})
The relations
\begin{align}
C_p(R|\specW(\specQ))&=
1- \lim_{\gamma \downarrow 0} {H}_p(\log d -R+\gamma|\specQ)
\Label{6-11-1}\\
C(\epsilon|\specW(\specQ))&=
\log d - H(1-\epsilon| \specQ)
\Label{6-11-2}
\end{align}
hold for $0\le \epsilon < 1$.
\end{thm}
This theorem and (\ref{6-11-11}) imply (\ref{6-11-4}).

\begin{rem}\rm
Verd\'{u} \& Han proved (\ref{6-11-2}) in the case of $\epsilon=0$ at (7.2) in \cite{Verdu-Han}.
Other cases are proven at the first time in this paper.
\end{rem}
Similar to Theorem \ref{thm31}, the following formulas for the second-order coding rate hold
for general additive channels.
\begin{thm}\Label{thm41}
The relations
\begin{align}
C_p(R_2,R_1|\specW)&=
1- \lim_{\gamma \downarrow 0} {H}_p(-R_2+\gamma,\log d - R_1|\specQ)
\Label{6-11-3}\\
C(\epsilon,R_1|\specW)&=
-H(1-\epsilon, \log d - R_1|\specQ)
\Label{6-11-4}
\end{align}
hold for $0\le \epsilon < 1$.
\end{thm}
Hence, we obtain Theorem \ref{thm11} from
(\ref{6-11-12}) and (\ref{6-11-13}).

Now, using Theorems \ref{thm3} and \ref{thm4}, we prove Theorems \ref{thm31} and \ref{thm41}.
Since $W^n_x(y)=Q^n(y-x)$, we have
\begin{align}
&{I}(\epsilon|\specP,\specW)
=
\left. \epliminf_{n \to \infty} 
\frac{1}{n}\log \frac{W^n_x(y)}{W^n_{P^n}(y)}\right|_{\rP_{P^n,W^n}}\nonumber \\
\le &
\left. \epliminf_{n \to \infty} 
\frac{1}{n}\log W^n_x(y)\right|_{\rP_{P^n,W^n}}
+
\left. \plimsup_{n \to \infty} 
\frac{-1}{n}\log {W^n_{P^n}(y)}\right|_{\rP_{P^n,W^n}}\Label{6-11-11} \\
\le &
\left. \epliminf_{n \to \infty} 
\frac{1}{n}\log Q^n(x)\right|_{Q^n}+\log d \nonumber \\
=& \log d - H(1-\epsilon| \specQ),\Label{6-11-10}
\end{align}
where 
(\ref{6-11-11}) and (\ref{6-11-10}) follow from 
(\ref{6-11-8}) and (\ref{6-11-9}), respectively.
Since the equality holds when $P^n$ is the uniform distribution,
we obtain
\begin{align*}
\sup_{\specP}I(\epsilon|\specP,\specW)
=\log d - H(1-\epsilon| \specQ),
\end{align*}
which implies (\ref{6-11-2}).
Similarly, we can show (\ref{6-11-4}).

Since $\plimsup_{n \to \infty} 
\frac{-1}{n}\log {W^n_{P^n}(y)}|_{W^n_{P^n}} \le d$,
we have
\begin{align*}
& \limsup_{n \to \infty}
\sum_{x \in {\cal X}_n} P^n(x) W^n_x
\left\{
\frac{1}{n}\log \frac{W^n_x(y)}{W^n_{P^n}(y)}
< R
\right\} \\
\ge &
\limsup_{n \to \infty}
\sum_{x \in {\cal X}_n} P^n(x) W^n_x
\left\{
\frac{1}{n}\log W^n_x(y)
+\log d < R
\right\} \\
= &
\limsup_{n \to \infty}
Q^n
\left\{
\frac{1}{n}\log Q^n(x)
< R-\log d 
\right\} \\
= &
1- \liminf_{n \to \infty}
Q^n
\left\{
\frac{-1}{n}\log Q^n(x)
< \log d -R 
\right\} ,
\end{align*}
which implies that
\begin{align*}
{I}_p(R|\specP,\specW)
\ge 
1- {H}_p(\log d -R|\specQ).
\end{align*}
Thus, we obtain (\ref{6-11-1}).
Similarly, we obtain (\ref{6-11-3}).

\begin{rem}\rm
When the sets 
${\cal X}_n$ and ${\cal Y}_n$ are given as general probability spaces with general $\sigma$-fields
$\sigma({\cal X}_n)$ and $\sigma({\cal Y}_n)$,
the above formulation can be extended with the following definition.
The $n$-th channel $W^n$ is given by the real-valued function from 
${\cal X}_n$ and $\sigma({\cal Y}_n)$ satisfying the following conditions;
(i) For any $x \in {\cal X}_n$, $W^n_x$ is a probability measure on ${\cal Y}_n$,
(ii) For any $F \in \sigma({\cal Y}_n)$, $W^n_{\cdot} (F)$ is a measurable function on ${\cal X}_n$.
$\specP$ and 
$\specQ$ take values in sequence of probability measures on ${\cal X}_n$ and ${\cal Y}_n$, respectively.
Then, 
the summands   
$\sum_{x\in {\cal X}_n} P^n(x)$ and 
$\sum_{y\in {\cal Y}_n} W^n_x(y)$ are replaced by 
$\int_{{\cal X}_n} P^n( d x)$ and 
$\int_{{\cal Y}_n} W^n_x(d y)$, respectively.
For any distribution $Q$ on ${\cal Y}_n$,
the function 
$\frac{W^n_x(y)}{Q(y)}$ is 
replaced by the inverse of Radon-Nikodym derivative 
$\frac{d Q}{d W^n_x}(y)$ of $Q$ with respect to $W^n_x$.
In the above definitions, 
$\inf_{\specP}$, $\sup_{\specP}$, $\inf_{\specQ}$, and $\sup_{\specQ}$ 
are given as the infimum and supremum among all sequences of probability measures on 
$\{{\cal X}_n\}_{n=1}^{\infty}$ 
and $\{{\cal Y}_n\}_{n=1}^{\infty}$.
The following proof is also valid in this extension.
\end{rem}

\section{Proof of the general formulas for the second-order coding rate}\Label{s9}
In this section, we prove Theorems \ref{thm3} and \ref{thm4}.
That is, for the reader's convenience, we present a proof for the first-order coding rate, as well as that for the second-order coding rate.
\subsection{Direct Part}\Label{s9a}
We prove the direct part, i.e., the inequalities
\begin{align}
C_p(R|\specW) & \le
\inf_{\specP}\lim_{\gamma \downarrow 0}{I}_p(R-\gamma|\specP,\specW) 
\Label{11-30-9}
\\
C(\epsilon|\specW) & \ge
\sup_{\specP}I(\epsilon|\specP,\specW)\Label{11-30-10}
\\
C_p(R_2,R_1|\specW)& \le
\inf_{\specP}\lim_{\gamma \downarrow 0}{I}_p(R_2-\gamma,R_1|\specP,\specW) \Label{11-30-11}
\\
C(\epsilon,R_1|\specW) &\ge
\sup_{\specP}{I}(\epsilon,R_1|\specP,\specW).\Label{11-30-12}
\end{align}
For arbitrary $R$, using the random coding method, we show that there exists a sequence of codes $\{\Phi_n\}$ such that $\frac{1}{n} \log |\Phi_n|\to R$ and 
$\limsup_{n \to \infty} P_{e,W^n}(\Phi_n) \le I_p(R|\specP,\specW)$.
This method is essentially the same as Verd\'{u} \& Han's method \cite{Verdu-Han}.

First, we set the size of $\Phi_{n,Z,R}$ to be $N_n=e^{nR - n^{\beta/2}}$ with the random variable $Z$. We generate the encoder $\phi_Z$, in which
$x \in {\cal X}^n$ is chosen as $\phi_Z(i)$ with the probability $P(x)$.
Here, the choice of $\phi_Z(i)$ is independent of the choice of other $\phi_Z(j)$.
The decoder $\{{\cal D}_{i,Z}\}_{i=1}^{N_n}$ is chosen 
by the following inductive method:
\begin{align*}
{\cal D}_{i,Z,R}\defeq
\left\{
\frac{1}{n}\log \frac{W^n_{\phi_Z(i)}(y)}{W^n_{P^n}(y)}
> R \right\}
\setminus \left(\bigcup_{j=1}^{i-1}{\cal D}_{j,Z}\right).
\end{align*}
Thus, 
the average error probability
is evaluated as
\begin{align*}
& E_Z P_{e,W^n}(\Phi_{n,Z,R})
\le
E_Z 
\frac{1}{N_n}
\sum_{i=1}^{N_n}
W^n_{\phi_Z(i)}\left(
\left\{
\frac{1}{n}\log \frac{W^n_{\phi_Z(i)}(y)}{W^n_{P^n}(y)}
> R \right\}^c
\bigcup
\left(\bigcup_{j=1}^{i-1}
\left\{
\frac{1}{n}\log \frac{W^n_{\phi_Z(j)}(y)}{W^n_{P^n}(y)}
> R \right\}\right)
\right) \\
\le &
E_Z 
\frac{1}{N_n}
\sum_{i=1}^{N_n}
W^n_{\phi_Z(i)}
\left\{
\frac{1}{n}\log \frac{W^n_{\phi_Z(i)}(y)}{W^n_{P^n}(y)}
< R \right\}
+
E_Z 
\frac{1}{N_n}
\sum_{i=1}^{N_n}
\sum_{j=1}^{i-1}
W^n_{\phi_Z(i)}
\left\{
\frac{1}{n}\log \frac{W^n_{\phi_Z(j)}(y)}{W^n_{P^n}(y)}
\ge R \right\}
\\
=&
\sum_{x} P^n(x) 
W^n_{x}
\left\{
\frac{1}{n}\log \frac{W^n_{x}(y)}{W^n_{P^n}(y)}
\le R \right\}
+
\frac{1}{N_n}
\sum_{i=1}^{N_n}
\sum_{j=1}^{i-1}
E_Z 
(E_Z 
W^n_{\phi_Z(i)})
\left\{
\frac{1}{n}\log \frac{W^n_{\phi_Z(j)}(y)}{W^n_{P^n}(y)}
\ge R \right\}.
\end{align*}
The second term is evaluated as
\begin{align*}
&
\frac{1}{N_n}
\sum_{i=1}^{N_n}
\sum_{j=1}^{i-1}
E_Z 
(E_Z 
W^n_{\phi_Z(i)})
\left\{
\frac{1}{n}\log \frac{W^n_{\phi_Z(j)}(y)}{W^n_{P^n}(y)}
\ge R \right\}
 \\
= &
\frac{1}{N_n}
\frac{N_n(N_n-1)}{2}
\sum_{x} P(x) 
W^n_{P}
\left\{
\frac{1}{n}\log \frac{W^n_{x}(y)}{W^n_{P^n}(y)}
\ge R \right\}\\
=& \frac{N_n-1}{2}
\sum_{x} P(x) 
W^n_{P}
\{W^n_{x}(y) e^{-nR} \ge W^n_{P^n}(y)\}
\\
\le &
\frac{N_n}{2} e^{-nR} 
\le \frac{e^{-n^{\beta/2}}}{2}
\to 0.
\end{align*}
Since
$\liminf_{n \to \infty} \sum_{x} P^n(x) 
W^n_{x}
\left\{
\frac{1}{n}\log \frac{W^n_{x}(y)}{W^n_{P^n}(y)}
\le R \right\}
=I_p(R|\specP,\specW)$,
(\ref{6-11-8}) implies that
$\liminf_{n \to \infty}
E_Z P_{e,W^n}(\Phi_{n,Z})
\le I_p(R|\specP,\specW)$.
Thus, the convergence $\frac{1}{n}\log |N_n| \to R$ implies
the inequality $C_p(R|\specW) \le \inf_{\specP}I_p(R|\specP,\specW) $.

Next, in order to prove (\ref{11-30-9}), 
for any sequence $\specP$,
we construct a code $\Phi_n$ such that $
\limsup_{n \to \infty}P_{e,W^n}(\Phi_n) \le \lim_{\gamma \downarrow 0}I_p(R_0-\gamma |\specP,\specW)$.
For any $k$, we choose the integer $N_k$ such that
$
E_Z P_{e,W^n}(\Phi_{n,Z,R_0-1/k})\le I_p(R_0-1/k |\specP,\specW) + 1/k 
$ for $\forall n \ge N_k$.
Then, for any $n$,
we choose $k(n)$ to be the maximum $k$ satisfying $n \ge N_k$. 
Then, $k(n) \to \infty$ as $n \to \infty$.
Thus,
$E_Z \Phi_{n,Z,R_0-1/k(n)}$ goes to $\lim_{\gamma \downarrow 0}I_p(R_0-\gamma |\specP,\specW)$,
and $\frac{1}{n}\log |\Phi_{n,Z,R_0-1/k(n)}|$ goes to $R_0$.
Hence, we obtain the inequality 
$C_p(R|\specW) \le \inf_{\specP}\lim_{\gamma \downarrow 0}I_p(R_0-\gamma|\specP,\specW) $, i.e., (\ref{11-30-9}).

For proving (\ref{11-30-11}),
we choose $N_n=e^{n R_1+n^{\beta}R_2 - n^{\beta/2}}$. 
Substituting $n R_1+n^{\beta} R_2$ into $n R$ in the above discussion, 
we denote the code $\Phi_{n,Z,R}$ by $\Phi_{n,Z,R_1,R_2}$.
Then,
\begin{align*}
E_Z P_{e,W^n}(\Phi_{n,Z,R_1,R_2})
\le
\sum_{x} P^n(x) 
W^n_{x}
\left\{
\frac{1}{n^{\beta}}
\left(\log \frac{W^n_{\phi_Z(i)}(y)}{W^n_{P^n}(y)}-nR_1\right)
< R_2  \right\} 
+
\frac{N_n}{2} e^{-(n R_1+ n^{\beta} R_2)} .
\end{align*}
Since $\frac{N_n}{2} e^{-(n R_1+ n^{\beta} R_2)}
\le \frac{e^{-n^{\beta/2}}}{2}
\to 0$ and
$\frac{1}{n^{\beta}}\log \frac{|N_n|}{e^{nR_1}} \to R_2$,
we obtain the inequality $C_p(R_2,R_1|\specW) \le \inf_{\specP} I_p(R_2,R_1|\specP,\specW) $.

For any $k$, we choose the integer $N_k$ such that
$E_Z P_{e,W^n}(\Phi_{n,Z,R_1,R_2-1/k})\le I_p(R_2-1/k,R_1 |\specP,\specW) + 1/k 
$ for $\forall n \ge N_k$.
Then, defining $k(n)$ similarly, we obtain
$E_Z \Phi_{n,Z,R_1,R_2-1/k(n)}\to \lim_{\gamma \downarrow 0}I_p(R_2-\gamma,R_1 |\specP,\specW)$,
and $\frac{1}{n^\beta}\log \frac{|\Phi_{n,Z,R_1,R_2-1/k(n)}|}{e^{n R_1}}\to R_2$.
Hence, we obtain the inequality 
$C_p(R_2,R_1|\specW) \le \inf_{\specP} \lim_{\gamma \downarrow 0}I_p(R_2-\gamma,R_1|\specP,\specW) $, i.e., 
(\ref{11-30-11}).

For an arbitrary number $R < \sup_{\specP}I(\epsilon|\specP,\specW)$, there exists a sequence of input distributions $\specP$ such that $I_p(R|\specP,\specW)\le \epsilon$. Therefore, the inequality (\ref{11-30-10}) holds. Similarly, we can show the inequality (\ref{11-30-12}).

\subsection{Converse part}
Next, we prove the converse part, i.e.,
\begin{align}
C_p(R|\specW) & \ge
\inf_{\specP}
\sup_{\specQ}\lim_{\gamma \downarrow 0}{J}_p(R-\gamma|\specP,\specQ,\specW) 
\Label{11-30-13}
\\
C(\epsilon|\specW) & \le
\sup_{\specP}
\inf_{\specQ}J(\epsilon|\specP,\specQ,\specW)\Label{11-30-14}
\\
C_p(R_2,R_1|\specW)& \ge
\inf_{\specP}
\sup_{\specQ}\lim_{\gamma \downarrow 0}{J}_p(R_2-\gamma,R_1|\specP,\specQ,\specW) \Label{11-30-15}
\\
C(\epsilon,R_1|\specW) &\le
\sup_{\specP}
\inf_{\specQ}J(\epsilon,R_1|\specP,\specQ,\specW),\Label{11-30-16}
\end{align}
which complete our proof, because the other inequalities
\begin{align*}
\inf_{\specP}\lim_{\gamma \downarrow 0}{I}_p(R-\gamma|\specP,\specW) 
& \le \inf_{\specP}
\sup_{\specQ}\lim_{\gamma \downarrow 0} {J}_p(R-\gamma|\specP,\specQ,\specW) \\
\sup_{\specP}I(\epsilon|\specP,\specW)
& \ge 
\sup_{\specP}
\inf_{\specQ}J(\epsilon|\specP,\specQ,\specW)\\
\inf_{\specP}\lim_{\gamma \downarrow 0} {I}_p(R_2-\gamma,R_1|\specP,\specW) 
& \le
\inf_{\specP}
\sup_{\specQ} \lim_{\gamma \downarrow 0} J_p(R_2-\gamma ,R_1|\specP,\specQ,\specW) \\
\sup_{\specP}I(\epsilon,R_1|\specP,\specW)
&\ge
\sup_{\specP}
\inf_{\specQ}J(\epsilon,R_1|\specP,\specQ,\specW)
\end{align*}
are trivial based on their definitions.
In the converse part, we essentially employ Hayashi-Nagaoka's\cite{H-N} method.
We choose an arbitrary sequence of codes $\{\Phi_n\}_{n=1}^{\infty}$.
Let $R$ be $\liminf_{n \to \infty} \frac{1}{n}\log |\Phi_n|$.
Assume that the code $\Phi_n$ consists of the triplet 
$(N_n, \phi, \{{\cal D}_i\}_{i=1}^{N_n})$.
Then, for any sequence of output distributions $\specQ=\{Q^n\}_{n=1}^{\infty}$
and any real $\gamma >0$,
the inequality 
\begin{align}
P_{e,W^n}(\Phi_n) \ge 
\sum_{x \in {\cal X}^n }
P_{\Phi_n}(x) W^n_{x}
\left\{
\frac{1}{n}\log \frac{W^n_{x}(y)}{Q^n(y)}
< R-\gamma \right\}
-  \frac{e^{n (R-\gamma)}}{N_n} 
\Label{11-30-17}
\end{align}
holds,
where $P_{\Phi_n}$ is the empirical distribution for the $|\Phi_n|$ points
$(\phi(1), \ldots, \phi(N_n))$.

Since $\frac{e^{n(R-\gamma)}}{N_n} \to 0$, the relation $\liminf_{n \to \infty} P_{e,W^n}(\Phi_n) \ge 
{J}_p(R-\gamma|\specPd,\specQ,\specW) $
holds for any $\specQ$, where $\specPd=\{P_{\Phi_n}\}$.
Thus, 
$\liminf_{n \to \infty} P_{e,W^n}(\Phi_n) \ge 
\sup_{\specQ}\lim_{\gamma \downarrow 0}{J}_p(R-\gamma|\specPd,\specQ,\specW) $.
Therefore, 
$\liminf_{n \to \infty} P_{e,W^n}(\Phi_n) \ge 
\inf_{\specP}\sup_{\specQ}\lim_{\gamma \downarrow 0}{J}_p(R-\gamma|\specPd,\specQ,\specW) $,
which implies (\ref{11-30-13}).

Now, assume that $\limsup_{n \to \infty}P_{e,W^n}(\Phi_n)=\epsilon$.
Since $\frac{e^{n(R-\gamma)}}{N_n} \to 0$, (\ref{11-30-17}) implies that
$R-\gamma \le J(\epsilon|\specP,\specQ,\specW)$.
Thus, $R-\gamma \le \sup_{\specP} \inf_{\specQ}J(\epsilon|\specP,\specQ,\specW)$, which implies 
Since $\gamma$ is an arbitrary positive real number, 
$R \le \sup_{\specP} \inf_{\specQ}J(\epsilon|\specP,\specQ,\specW)$, which implies (\ref{11-30-14}).

Next, consider the case in which $\liminf_{n \to \infty} 
\frac{1}{n^{\beta}}\log \frac{|\Phi_n|}{e^{nR_1}}= R_2$.
Replacing $R-\gamma$ by $R_1 +n^{\beta-1}(R_2-\gamma)$ in (\ref{11-30-17}),
we obtain $\frac{e^{nR_1 +n^{\beta}(R_2-\gamma)}}{N_n} \to 0$.
Thus, 
$\liminf_{n \to \infty} P_{e,W^n}(\Phi_n) \ge 
\inf_{\specP}\sup_{\specQ}\lim_{\gamma \downarrow 0}
J_p(R_2-\gamma,R_1|\specP,\specQ,\specW) $,
which implies (\ref{11-30-15}).
replacing $R_1 + R_2 n^{\beta-1}$ into $R-\gamma$ in (\ref{11-30-17}),
similar to (\ref{11-30-14}),
we can show (\ref{11-30-16}).

The inequality (\ref{11-30-17}) is shown as follows.
We focus on the inequalities:
\begin{align*}
& W^n_{\phi(i)}({\cal D}_i) - e^{n R' } Q^n({\cal D}_i) \\
\le & 
W^n_{\phi(i)}(\{ W^n_{\phi(i)}(y) - e^{n R' } Q^n(y) \ge 0 \})
- e^{n R' } Q^n(\{ W^n_{\phi(i)}(y) - e^{n R' } Q^n(y) \ge 0 \}) \\
\le &
W^n_{\phi(i)}(\{ W^n_{\phi(i)}(y) - e^{n R' } Q^n(y) \ge 0 \}) \\
= & 
W^n_{\phi(i)}
\left\{
\frac{1}{n}\log \frac{W^n_{\phi(i)}(y)}{Q^n(y)}
\ge R' \right\},
\end{align*}
where the first inequality follows from the fact that
any two distributions $P$ and $Q$ and any positive constant $a$ satisfy
$\max_{{\cal D}} [P({\cal D})- a Q ({\cal D})]
=P\{P(\omega)-a Q(\omega) \ge 0\}
-a Q \{P(\omega)-a Q(\omega) \ge 0\}$.

Thus, 
\begin{align*}
& 1- P_{e,W^n}(\Phi_n) 
= \frac{1}{N_n}\sum_{i=1}^{N_n} W^n_{\phi(i)}({\cal D}_i) \\
\le &
\frac{1}{N_n}\sum_{i=1}^{N_n}
e^{n R' } Q^n({\cal D}_i)+
W^n_{\phi(i)}
\left\{
\frac{1}{n}\log \frac{W^n_{\phi(i)}(y)}{Q^n(y)}
\ge R' \right\}
\\
= &
\frac{e^{n R' }}{N_n}
+
1-\sum_{x \in {\cal X}^n }
P_{\Phi_n}(x) W^n_{x}
\left\{
\frac{1}{n}\log \frac{W^n_{x}(y)}{Q^n(y)}
< R' \right\},
\end{align*}
which implies (\ref{11-30-17}).

\section{Proof of the stationary memoryless case}\Label{s10}
\subsection{Proof of Theorem \ref{thm2}}\Label{s10a}
In this subsection, using Theorem \ref{thm4},
we prove Theorem \ref{thm2} when the cardinality $|{\cal X}|$ is finite. 
For this purpose, we show the following relations
in the stationary discrete memoryless case, i.e., the case in which
$W_x^n(y)= W^{\times n}_x(y) \defeq\prod_{i=1}^nW_{x_i}(y_i)$ for 
$x=(x_1, \ldots, x_n)$ and $y=(y_1, \ldots, y_n)$.
In this section, abbreviating $C^{\dm}_W$ as $C$, we will prove that
\begin{align}
\inf_{\specP} \lim_{\gamma \downarrow 0}  {I}_p(R_2-\gamma,C|\specP,\specW)
\le
\left\{ 
\begin{array}{cc}
\erf(R_2/\sqrt{V_W^+}) & R_2 \ge 0 \\
\erf(R_2/\sqrt{V_W^-}) & R_2 < 0.
\end{array}
\right.
\Label{11-1-a}
\end{align}
and 
\begin{align}
\inf_{\specP}\sup_{\specQ}
\lim_{\gamma \downarrow 0} {J}_p(R_2-\gamma,C|\specP,\specQ,\specW)
\ge
\left\{ 
\begin{array}{cc}
\erf(R_2/\sqrt{V_W^+}) & R_2 \ge 0 \\
\erf(R_2/\sqrt{V_W^-}) & R_2 < 0.
\end{array}
\right.
\Label{11-1-b}
\end{align}
Showing both inequalities and using Theorem \ref{thm4}, we obtain
\begin{align}
C_p(R_2,R_1|\specW)
=&
\left\{ 
\begin{array}{cc}
\erf(R_2/\sqrt{V_W^+}) & R_2 \ge 0 \\
\erf(R_2/\sqrt{V_W^-}) & R_2 < 0.
\end{array}
\right.
\Label{12-10-10}
\end{align}
Since the rhs of (\ref{12-10-10}) is continuous with respect to $\epsilon$,
(\ref{12-10-10}) implies that
\begin{align*}
C(\epsilon,R_1|\specW)
=
\left\{ 
\begin{array}{cc}
\sqrt{V_W^+} \erf^{-1}(\epsilon) & \epsilon \ge 1/2\\
\sqrt{V_W^-} \erf^{-1}(\epsilon) & \epsilon < 1/2.
\end{array}
\right.
\end{align*}
That is, we can show Theorem \ref{thm2}.

In fact, when $\specP$ is the i.i.d. of $P_{M+}$ or $P_{M-}$,
$I(\epsilon,C|\specP,\specW)$
is equal to $\sqrt{V_W^+} F^{-1}(\epsilon)$ or 
$\sqrt{V_W^-} F^{-1}(\epsilon)$.
Thus, (\ref{11-1-a}) holds. 
Therefore, the achievability part (the direct part) of Theorem \ref{thm2}
hold.
Therefore, it is sufficient to prove the converse part (\ref{11-1-b}).

\begin{figure}[htbp]
\begin{center}
\scalebox{1.0}{\includegraphics[scale=0.5]{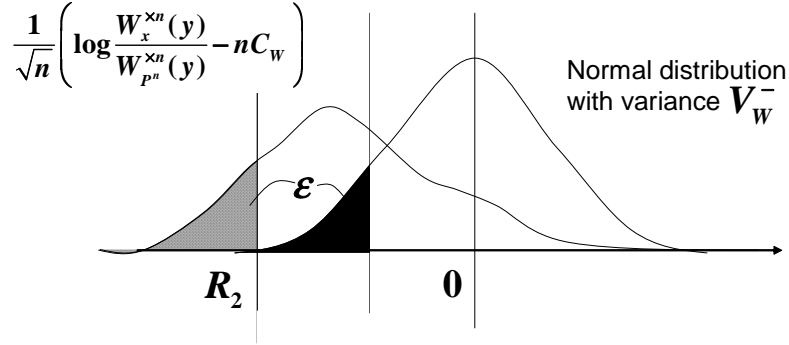}}
\end{center}
\caption{
Limiting behavior of $\frac{1}{\sqrt{n}}\left(\log \frac{W^{\times n}_x(y)}{W^{\times n}_{P^n}(y)}-nC\right)$ and the Gaussian distribution with the variance $V_W^-$
}
\Label{pic3}
\end{figure}%

We focus on the set $T_n$ of empirical distributions with $n$ outcomes.
Its cardinality $|T_n|$ is evaluated as
$|T_n|\le (n+1)^{|{\cal X}|}$.
In this proof, we use the distribution 
\begin{align*}
Q^{n}_U\defeq \sum_{P \in T_n}\frac{1}{|T_n|+1}(W_P)^{\times n}
+ \frac{1}{|T_n|+1}Q_{M}^{\times n}
\end{align*}
and the sets 
\begin{align*}
{\cal V}_{\epsilon} &\defeq
\{ P | I(P,W) \ge C + \epsilon\} \\
\Omega_n &\defeq \left\{x\in {\cal X}^n \left|  
\ep(x) \in 
{\cal V}_{\epsilon} \right.\right\} ,
\end{align*}
where 
$\ep(x)$ is the empirical distribution of $x\in {\cal X}^n$.

Since 
$Q^{n}_U (y) \ge
\frac{1}{|T_n|+1}(W_{\ep(x)})^{\times n}(y)$
and 
$Q^{n}_U(y) \ge
\frac{1}{|T_n|+1}Q_{M}^{\times n}(y)$,
\begin{align*}
& {\rm P}_{P^n,W^{\times n}}
\left\{
\frac{1}{\sqrt{n}}\left(\log \frac{W^{\times n}_x(y)}{Q^{n}_U(y)}
-nC\right)
\le R
\right\} \\
=&
\sum_{x \in \Omega_n}P^n(x)
{\rm P}_{W^{\times n}_x}
\left\{
\frac{1}{\sqrt{n}}\left(\log \frac{W^{\times n}_x(y)}{Q^{n}_U(y)}
-nC\right)
\le R
\right\} \\
& +
\sum_{x \in \Omega_n^c}P^n(x)
{\rm P}_{W^{\times n}_x}
\left\{
\frac{1}{\sqrt{n}}\left(\log \frac{W^{\times n}_x(y)}{Q^{n}_U(y)}
-nC\right)
\le R
\right\} \\
\ge &
\sum_{x \in \Omega_n}P^n(x)
{\rm P}_{W^{\times n}_x}
\left\{
\frac{1}{\sqrt{n}}\left(\log \frac{W^{\times n}_x(y)}{(Q_{M})^{\times n}(y)}
+\log (|T_n|+1)-nC\right)
\le R
\right\} \\
& +
\sum_{x \in \Omega_n^c}P^n(x)
{\rm P}_{W^{\times n}_x}
\left\{
\frac{1}{\sqrt{n}}\left(\log \frac{W^{\times n}_x(y)}{(W_{\ep(x)})^{\times n}(y)}
+\log (|T_n|+1)-nC\right)
\le R
\right\}.
\end{align*}

When $x \in {\cal V}_{\epsilon}^c$,
\begin{align*}
{\rm V}_{W^{\times n}_x}
\frac{1}{\sqrt{n}}\left(\log \frac{W^{\times n}_x(y)}{(W_{\ep(x)})^{\times n}(y)}
-nC\right)
& =V_{\ep(x),W}< \max_{P}V_{P,W}\\
{\rm E}_{W^{\times n}_x}
\frac{1}{\sqrt{n}}\left(\log \frac{W^{\times n}_x(y)}{(W_{\ep(x)})^{\times n}(y)}
+\log (|T_n|+1)-nC\right)
&=
\frac{1}{\sqrt{n}}
\left( n I(\ep(x),W)+\log (|T_n|+1)-nC \right) \\
& \le 
\frac{\log (|T_n|+1)}{\sqrt{n}}
-\epsilon \sqrt{n}.
\end{align*}
Thus, Chebyshev inequality implies
\begin{align*}
& {\rm P}_{W^{\times n}_x}
\left\{
\frac{1}{\sqrt{n}}\left(\log \frac{W^{\times n}_x(y)}{(W_{\ep(x)})^{\times n}(y)}
+\log (|T_n|+1)-nC\right)
\le R 
\right\}\\
\ge & 
1- \frac{\max_{P}V_{P,W}}{R+\epsilon \sqrt{n}-\frac{\log (|T_n|+1)}{\sqrt{n}} }.
\end{align*}
Define the quantity $V_{P,W}'\defeq 
{\rm E}_{P} {\rm E}_{W_x} 
(\log \frac{W_x(y)}{Q_M(y)} -D(W_x\|Q_M))^2$.
When $x \in {\cal V}_{\epsilon}$,
since the random variable $\log \frac{W^{\times n}_x(y)}{(Q_{M})^{\times n}(y)}
=\sum_{i=1}^n \log \frac{W_{x_i}(y_i)}{(Q_{M})(y_i)}$
has the variance
$n V_{\ep(x),W}'$,
\begin{align*}
& {\rm P}_{W^{\times n}_x}
\left\{
\frac{1}{\sqrt{n}}\left(\log \frac{W^{\times n}_x(y)}{(Q_{M})^{\times n}(y)}
+\log (|T_n|+1)-nC\right)
\le R
\right\} \\
\ge &
 {\rm P}_{W^{\times n}_x}
\left\{
\frac{1}{\sqrt{n}}\left(
\log \frac{W^{\times n}_x(y)}{(Q_{M})^{\times n}(y)}
+\log (|T_n|+1)-nI(\ep(x),W) \right)
\le R
\right\} \\
\cong &
G\left(\frac{R}{\sqrt{V_{\ep(x),W}'}}\right)
\ge 
\min_{P \in {\cal V}_{\epsilon}}
G\left(\frac{R}{\sqrt{V_{P,W}'}}\right).
\end{align*}
Since the random variable $\log \frac{W^{\times n}_x(y)}{(Q_{M})^{\times n}(y)}
=\sum_{i=1}^n \log \frac{W_{x_i}(y_i)}{(Q_{M})(y_i)}$
is written as a combination of finite number of random variables 
$\{\log \frac{W_{x}(y)}{(Q_{M})(y)}\}_{x \in {\cal X}}$,
the above convergence is uniform.
That is, for any $\delta >0$, there exists $N>0$ such that
for $n \ge N$,
\begin{align*}
& {\rm P}_{W^{\times n}_x}
\left\{
\frac{1}{\sqrt{n}}\left(\log \frac{W^{\times n}_x(y)}{(Q_{M})^{\times n}(y)}
+\log (|T_n|+1)-nC\right)
\le R
\right\} \\
\ge &
\min_{P \in {\cal V}_{\epsilon}}
G\left(\frac{R}{\sqrt{V_{P,W}'}}\right) - \delta.
\end{align*}
Therefore,
\begin{align*}
& {\rm P}_{P^n,W^{\times n}}
\left\{
\frac{1}{\sqrt{n}}\left(\log \frac{W^{\times n}_x(y)}{Q^{n}_U(y)}
-nC\right)
\le R
\right\} \\
\ge &
P^n(\Omega_n)
(1- \frac{\max_{P}V_{P,W}}{R+\epsilon \sqrt{n} -\frac{\log (|T_n|+1)}{\sqrt{n}} })
+
P^n(\Omega_n^c)
\min_{P \in {\cal V}_{\epsilon}}
G\left(\frac{R}{\sqrt{V_{P,W}'}}\right)- \delta \\
\ge &
\min_{P \in {\cal V}_{\epsilon}}
G\left(\frac{R}{\sqrt{V_{P,W}'}}\right)- \delta ,
\end{align*}
where $\Omega_n^c$ is the complement of $\Omega_n$.

Thus,
\begin{align*}
& \limsup_{n \to \infty} {\rm P}_{P^n,W^{\times n}}
\left\{
\frac{1}{\sqrt{n}}\left(\log \frac{W^{\times n}_x(y)}{Q^{n}_U(y)}
-nC\right)
\le R
\right\} \\
\ge &
\min_{P \in {\cal V}_{\epsilon}}
G\left(\frac{R}{\sqrt{V_{P,W}}}\right)
-\delta.
\end{align*}
Since $\delta >0$ and $\epsilon > 0$ are arbitrary,
when $\specQ=\{Q^{n}_U\}$,
\begin{align*}
&J_p(R,C|\specP,\specQ,\specW ) \\
=& \limsup_{n \to \infty} {\rm P}_{P^n,W^{\times n}}
\left\{
\frac{1}{\sqrt{n}}\left(\log \frac{W^{\times n}_x(y)}{Q^{n}_U(y)}
-nC\right)
\le R
\right\} \\
\ge &
\min_{P \in {\cal V}}
G\left(\frac{R}{\sqrt{V_{P,W}}}\right)
=
\left\{ 
\begin{array}{cc}
\erf(R/\sqrt{V_W^+}) & R \ge 0 \\
\erf(R/\sqrt{V_W^-}) & R < 0.
\end{array}
\right. 
\end{align*}
which implies (\ref{11-1-b}) because of the continuity of the r.h.s.

\subsection{Proof of Theorem \ref{thm2b}}\Label{s10b}
In this subsection, using Theorem \ref{thm4b},
we prove Theorem \ref{thm2b} when the cardinality $|{\cal X}|$ is finite. 
For this purpose, we show the following relations
in the stationary discrete memoryless case, i.e., the case in which
$W_x^n(y)= W^{\times n}_x(y) \defeq\prod_{i=1}^nW_{x_i}(y_i)$ for 
$x=(x_1, \ldots, x_n)$ and $y=(y_1, \ldots, y_n)$,
and $c_n(x)=\sum_{i=1}^n c(x_i)$.
In this section, abbreviating $C^{\dm}_W$ as $C$, we will prove that
\begin{align}
\inf_{\specP:\supp(P_n)\subset {\cal X}_{n,c,K}}
\lim_{\gamma \downarrow 0} {I}_p(R_2-\gamma,R_1|\specP,\specW)
\le
\left\{ 
\begin{array}{cc}
\erf(R_2/\sqrt{V_{W,c,K}^+}) & R_2 \ge 0 \\
\erf(R_2/\sqrt{V_{W,c,K}^-}) & R_2 < 0.
\end{array}
\right.
\Label{11-1-ab}
\end{align}
and 
\begin{align}
\inf_{\specP:\supp(P_n)\subset {\cal X}_{n,c,K}}\sup_{\specQ} \lim_{\gamma \downarrow 0} {J}_p(R_2-\gamma,R_1|\specP,\specQ,\specW)
\ge
\left\{ 
\begin{array}{cc}
\erf(R_2/\sqrt{V_{W,c,K}^+}) & R_2 \ge 0 \\
\erf(R_2/\sqrt{V_{W,c,K}^-}) & R_2 < 0.
\end{array}
\right.
\Label{11-1-bb}
\end{align}
Showing both inequalities and using Theorem \ref{thm4b}, we obtain
\begin{align}
C_p(R_2,R_1|\specW,\specc,K)
=&
\left\{ 
\begin{array}{cc}
\erf(R_2/\sqrt{V_{W,c,K}^+}) & R_2 \ge 0 \\
\erf(R_2/\sqrt{V_{W,c,K}^-}) & R_2 < 0.
\end{array}
\right.
\Label{12-10-10b}
\end{align}
Since the rhs of (\ref{12-10-10b}) is continuous with respect to $\epsilon$,
(\ref{12-10-10b}) implies that
\begin{align*}
C(\epsilon,R_1|\specW,\specc,K)
=
\left\{ 
\begin{array}{cc}
\sqrt{V_{W,c,K}^+} \erf^{-1}(\epsilon) & \epsilon \ge 1/2\\
\sqrt{V_{W,c,K}^-} \erf^{-1}(\epsilon) & \epsilon < 1/2.
\end{array}
\right.
\end{align*}
That is, we can show Theorem \ref{thm2b}.

The inequality (\ref{11-1-bb}) can be proven in the same way as (\ref{11-1-b})
by replacing $T_n$ and $Q_M$ by the set of empirical distributions
$T_{n,c,K}\defeq \{P\in T_n| {\rm E}_P c(x) \le K\}$.
and $Q_{M,c,K}$.
Therefore, the converse part of Theorem \ref{thm2b}
hold.
Therefore, it is sufficient to prove the direct part (\ref{11-1-ab}).

For any distribution $P$ satisfying ${\rm E}_P c(x)\le K$,
we choose the closet empirical distribution $P_n \in T_{n,c,K}$.
Let $\specP=\{P^n\}$ be the uniform distributions on the set $T_{P_n}\defeq \{x \in {\cal X}^n| \ep(x) = P_n\}$.
It is sufficient to show that
\begin{align}
I_p(R,C|\specP,\specW ) 
\le
\erf(R/\sqrt{V_{P,W}}) .\Label{8-7-1}
\end{align}
Since
\begin{align}
P^n(x) \le |T_n|(P_n)^{\times n}(x),\Label{8-20-1}
\end{align}
we have
\begin{align}
&I_p(R,C|\specP,\specW ) \nonumber \\
=& \limsup_{n\to \infty} {\rm P}_{P^n,W^{\times n}}
\left\{
\frac{1}{\sqrt{n}}\left(\log \frac
{W^{\times n}_x(y)}{W^{\times n}_{P^n}(y)}
-nC\right)
\le R
\right\} \nonumber \\
\le &
\limsup_{n\to \infty} {\rm P}_{P^n,W^{\times n}}
\left\{
\frac{1}{\sqrt{n}}\left(\log \frac
{W^{\times n}_x(y)}{(W_{P_n})^{\times n}(y)}
-\log |T_n|
-nC\right)
\le R
\right\} \nonumber \\
\le &
G\left(\frac{R}{\sqrt{V_{P,W}}}\right),\Label{8-20-2}
\end{align}
which implies (\ref{8-7-1}).

In order to prove (\ref{8-7-1}) without condition $|{\cal X}|< \infty$,
we choose a sequence of input distributions $\{P^{(k)}_{\pm}\}_{k=1}^{\infty}$ with finite supports such that
\begin{align*}
P^{(k)} & \in T_{n,c,K} \\
I(P^{(k)}_{\pm},W) &\to \max_{P:{\rm E}_P c(x)\le K} I(P,W) \\ 
V_{P^{(k)}_{\pm},W} &\to V_{W,c,K}^{\pm}.
\end{align*}
Choose the distribution $P^n$ as
the uniform distributions on the set $T_{P^{(n^{\frac{1}{4}})}}$.
Then, in stead of (\ref{8-20-1}), the relation
\begin{align*}
P^n(x) \le (n+1)^{n^{\frac{1}{4}}} (P^{(n^{\frac{1}{4}})})^{\times n}(x)
\end{align*}
holds.
Since $\frac{1}{\sqrt{n}}\log (n+1)^{n^{\frac{1}{4}}}$ goes to zero,
the same discussion as (\ref{8-20-2}) yields (\ref{8-7-1}).

\subsection{Proof of Theorem \ref{thm2c}}
As is shown in Subsection \ref{s10b},
we obtain the direct part, i.e., 
\begin{align*}
C^{G}_p(a,C^G_{N,S}|N,S)
\le \erf(a/\sqrt{V_{P_M,W}}) .
\end{align*}
Hence, when $c_n(x)=\sum_{i=1}^n x_i^2$,
it is sufficient to prove
\begin{align}
\inf_{\specP:\supp(P_n)\subset {\cal X}_{n,c,S}}\sup_{\specQ} \lim_{\gamma \downarrow 0} {J}_p(R_2-\gamma,R_1|\specP,\specQ,\specW)
\ge
 \erf(a/\sqrt{V_{P_M,W}}) .
\Label{11-1-bc}
\end{align}
In the following discussion, we use the distribution 
\begin{align*}
Q^{n}_U & \defeq
\frac{1}{2}(W_{P_M})^{\times n}
+ 
\frac{1}{2}(W_{P_{M,\epsilon}})^{\times n} \\
P_{M,\epsilon} & \defeq \frac{1}{\sqrt{2 \pi (S-\epsilon)}}
e^{-\frac{x^2}{2(S-\epsilon)}}
\end{align*}
and the sets
\begin{align*}
{\cal V}_{\epsilon} &\defeq
\{ P | {\rm E}_{P} x^2 \le S - \epsilon\} \\
\Omega_n &\defeq \left\{x\in {\cal X}^n \left|  
\ep(x) \in 
{\cal V}_{\epsilon} \right.\right\} .
\end{align*}
We obtain
\begin{align*}
& {\rm P}_{P^n,W^{\times n}}
\left\{
\frac{1}{\sqrt{n}}\left(\log \frac{W^{\times n}_x(y)}{Q^{n}_U(y)}
-nC\right)
\le R
\right\} \\
=&
\sum_{x \in \Omega_n}P^n(x)
{\rm P}_{W^{\times n}_x}
\left\{
\frac{1}{\sqrt{n}}\left(\log \frac{W^{\times n}_x(y)}{Q^{n}_U(y)}
-nC\right)
\le R
\right\} \\
& +
\sum_{x \in \Omega_n^c}P^n(x)
{\rm P}_{W^{\times n}_x}
\left\{
\frac{1}{\sqrt{n}}\left(\log \frac{W^{\times n}_x(y)}{Q^{n}_U(y)}
-nC\right)
\le R
\right\} \\
\ge &
\sum_{x \in \Omega_n}P^n(x)
{\rm P}_{W^{\times n}_x}
\left\{
\frac{1}{\sqrt{n}}\left(\log \frac{W^{\times n}_x(y)}{(W_{P_M})^{\times n}(y)}
+\log 2-nC\right)
\le R
\right\} \\
& +
\sum_{x \in \Omega_n^c}P^n(x)
{\rm P}_{W^{\times n}_x}
\left\{
\frac{1}{\sqrt{n}}\left(\log \frac{W^{\times n}_x(y)}{(W_{P_{M,\epsilon}})^{\times n}(y)}
+\log 2 -nC\right)
\le R
\right\}.
\end{align*}

When $x \in {\cal V}_{\epsilon}^c$,
the random variable 
$\frac{1}{\sqrt{n}}\left(\log \frac{W^{\times n}_x(y)}{(W_{P_{M,\epsilon}})^{\times n}(y)}
+\log 2-nC\right)$
has 
the expectation 
\par\indent $\frac{1}{\sqrt{n}}
\left(\frac{n}{2}\log (1+\frac{S-\epsilon }{N})
+\frac{\frac{\|x\|^2}{n N}-\frac{S-\epsilon }{N}}{2(1+\frac{S-\epsilon }{N})}
-\frac{n}{2}\log (1+\frac{S}{N})
+\log 2 \right) 
(\le 
\frac{\log 2}{\sqrt{n}}
-\frac{\sqrt{n}}{2}\log \frac{1+\frac{S}{N}}{1+\frac{S-\epsilon}{N}})$,
and the variance 
$\frac{\frac{(S-\epsilon)^2}{N^2} + 2 \frac{\|x\|^2}{n N}}{2(1+\frac{S-\epsilon}{N})^2}
(\le
\frac{\frac{(S-\epsilon)^2}{N^2} + 2 \frac{S-\epsilon}{N}}{2(1+\frac{S-\epsilon}{N})^2})$.
Thus, Chebyshev inequality implies
\begin{align*}
& {\rm P}_{W^{\times n}_x}
\left\{
\frac{1}{\sqrt{n}}\left(\log \frac{W^{\times n}_x(y)}{(W_{P_{M,\epsilon}})^{\times n}(y)}
+\log 2-nC\right)
\le R 
\right\}\\
\ge & 
1- \frac{\frac{\frac{(S-\epsilon)^2}{N^2} + 2 \frac{S-\epsilon}{n N}}{2(1+\frac{S-\epsilon}{N})^2}
}{R
+\frac{\sqrt{n}}{2}\log \frac{1+\frac{S}{N}}{1+\frac{S-\epsilon}{N}}
-\frac{\log 2}{\sqrt{n}}} \to 1.
\end{align*}
When $x \in {\cal V}_{\epsilon}$,
under the $n$-variable Gaussian distribution $W^{\times n}_x$, 
the random variable
$
\log \frac{W^{\times n}_x(y+x)}{(W_{P_M})^{\times n}(y+x)}$
is calculated to be
\begin{align*}
\frac{1}{2(1+\frac{S}{N})}
\left(-\frac{S\|y\|^2}{N^2}+ \frac{2x \cdot y}{N} +\frac{\|x\|^2}{N}\right)
\frac{n}{2}\log (1+\frac{S}{N}).
\end{align*}
The expectation is 
$\frac{\frac{\|x\|^2}{N}-n\frac{S}{N}}{2(1+\frac{S}{N})}
+\frac{n}{2}\log (1+\frac{S}{N})$, and
the variance is 
$\frac{2 n \frac{S^2}{N^2}+4\frac{\|x\|^2}{N}}{4(1+\frac{S}{N})^2}$.
The random variable 
\par\indent $
\frac{1}{\sqrt{n}}\left(\log \frac{W^{\times n}_x(y+x)}{(W_{P_M})^{\times n}(y+x)}
-
\frac{\frac{\|x\|^2}{N}-n\frac{S}{N}}{2(1+\frac{S}{N})}
-\frac{n}{2}\log (1+\frac{S}{N})
\right)
$ converges the normal distribution when $n$ goes to infinity.
Due to the property of Gaussian distribution,
this convergence is uniform when $\|x\|$ is bounded.
Hence,
\begin{align*}
& {\rm P}_{W^{\times n}_x}
\left\{
\frac{1}{\sqrt{n}}\left(
\log \frac{W^{\times n}_x(y)}{(W_{P_M})^{\times n}(y)}
+\log 2-nC\right)
\le R
\right\} \\
\ge &
 {\rm P}_{W^{\times n}_x}
\left\{
\frac{1}{\sqrt{n}}\left(
\log \frac{W^{\times n}_x(y)}{(W_{P_M})^{\times n}(y)}
+\log 2 - \frac{\frac{\|x\|^2}{N}-n\frac{S}{N}}{2(1+\frac{S}{N})}
-\frac{n}{2}\log (1+\frac{S}{N})
 \right)
\le R
\right\} \\
\cong &
G\left(\frac{R}{\sqrt{\frac{2 \frac{S^2}{N^2}+4\frac{\|x\|^2}{n N}}{4(1+\frac{S}{N})^2}}}
\right) \\
\ge &
\left\{
\begin{array}{cc}
G\left(\frac{R}{\sqrt{\frac{2 \frac{S^2}{N^2}+4\frac{S-\epsilon}{N}}{4(1+\frac{S}{N})^2}}}\right) & R \le 0\\
G\left(\frac{R}{\sqrt{\frac{2 \frac{S^2}{N^2}+4\frac{S}{N}}{4(1+\frac{S}{N})^2}}}\right) & R > 0.
\end{array}
\right.
\end{align*}
Therefore,
\begin{align*}
& \limsup_{n \to \infty} {\rm P}_{P^n,W^{\times n}}
\left\{
\frac{1}{\sqrt{n}}\left(\log \frac{W^{\times n}_x(y)}{Q^{n}_U(y)}
-nC\right)
\le R
\right\} \\
\ge &
\left\{
\begin{array}{cc}
G\left(\frac{R}{\sqrt{\frac{2 n \frac{S^2}{N^2}+4n\frac{S-\epsilon}{N}}{4(1+\frac{S}{N})^2}}}\right) & R \le 0\\
G\left(\frac{R}{\sqrt{\frac{2 n \frac{S^2}{N^2}+4n\frac{S}{N}}{4(1+\frac{S}{N})^2}}}\right) & R > 0
\end{array}
\right.
\end{align*}
Since $\epsilon > 0$ is arbitrary,
when $\specQ=\{Q^{n}_U\}$,
\begin{align*}
&J_p(R,C|\specP,\specQ,\specW ) \\
=& \limsup_{n \to \infty} {\rm P}_{P^n,W^{\times n}}
\left\{
\frac{1}{\sqrt{n}}\left(\log \frac{W^{\times n}_x(y)}{Q^{n}_U(y)}
-nC\right)
\le R
\right\} \\
\ge &
G\left(\frac{R}{\sqrt{\frac{2 \frac{S^2}{N^2}+4\frac{S}{N}}{4(1+\frac{S}{N})^2}}}\right) ,
\end{align*}
which implies (\ref{11-1-bc}).

\section{Concluding remarks and future study}\Label{s11}
We have obtained a general asymptotic formula for channel coding in the sense of the second-order coding rate. That is, it has been shown that the optimum second-order transmission rate with the error probability $\epsilon$ is characterized by the second-order asymptotic behavior of the logarithmic likelihood ratio between the conditional output distribution and the non-conditional output distribution. Using this result, we have derived this type of optimal transmission rate for 
the discrete memoryless case,
the discrete memoryless case with a cost constraint,
the additive Markovian case,
and the Gaussian channel case with an energy constraint.
The performance in the second-order coding rate is characterized by the average of the variance of the logarithmic likelihood ratio with the single letterized expression. 
When the input distribution producing the capacity is not unique, it is characterized by its minimum and its maximum. We give a typical example such that the minimum is different from the maximum. Furthermore, both quantities have been verified to satisfy the additivity.

The main results of the present study are as follows. While the application of the information spectrum method to the second-order coding rate was initiated by Hayashi \cite{H-sec}, his research indicated that there is no difficulty in extending general formulas to the second-order coding rate. Therefore, in the i.i.d. case, the second-order coding rate of the source coding and intrinsic randomness are solved by the central limit theorem. 
However, channel coding cannot been treated using the method of Hayashi\cite{H-sec} 
except for the additive noise case with no cost constraint
because the present problem contains the optimization concerning the input distribution in the non-additive noise case.
In the converse part, we have to treat the general sequence of input distributions. 
In order to resolve this difficulty, we have treated the logarithmic likelihood ratio between the conditional output distribution and the distribution $Q_U^n$, which is introduced in Subsection \ref{s10a}.

Furthermore, we can consider the quantum extension of our results. There is considerable difficulty concerning non-commutativity in this direction. In addition, the third-order coding rate is expected but appears difficult. The second order is the order $\sqrt{n}$, and it is not clear whether the third order is a constant order or the order $\log n$. This is an interesting problem for future study.

\section*{Acknowledgments}
This study was supported by MEXT through a Grant-in-Aid for Scientific Research on Priority Area "Deepening and Expansion of Statistical Mechanical Informatics (DEX-SMI)", No. 18079014. 
The author thanks Professor Tomohiko Uyematsu for informing the reference \cite{strassen}.

\appendix

For a given $R<0$, we prove (\ref{6-16-1}).
Since $\frac{d^2 \psi_P}{ds^2}(s) >0$, the function $\psi_P$ is convex.
Choosing $s_n$ such that
$C_W^{\dm}+\frac{R_2}{\sqrt{n}}=
-\frac{d \psi_P}{ds}(s_n)
=-\frac{d \psi_P}{ds}(0)
-\int_{0}^{s_n} \frac{d^2 \psi_P}{ds^2}(t) dt$,
we have the relation 
\begin{align}
\frac{R_2}{\sqrt{n}}=-\int_{0}^{s_n} \frac{d^2 \psi_P}{ds^2}(t) dt \Label{6-16-2}.
\end{align}
Then, the minimum of $C_W^{\dm}s + \frac{R_2}{\sqrt{n}} s+\psi_P(s)$
is attained when $s=s_n$.
Since $\frac{d^2 \psi_P}{ds^2}(s)$ is continuous and bounded,
$s_n$ approaches zero as $n$ goes to infinity.
More precisely, (\ref{6-16-2}) implies 
$R_2 = -\lim_{n\to \infty} \sqrt{n} \int_{0}^{s_n} \frac{d^2 \psi_P}{ds^2}(t) dt
= -\lim_{n\to \infty} (\sqrt{n} s_n) \frac{d^2 \psi_P}{ds^2}(0)$.
That is, 
$\lim_{n\to \infty} (\sqrt{n} s_n)= \frac{-R_2 }{\frac{d^2 \psi_P}{ds^2}(0)}$.
When the function $\epsilon(u)$ is chosen to be $\frac{d^2 \psi_P}{ds^2}(u) -\frac{d^2 \psi_P}{ds^2}(0) $,
$\epsilon(u)$ approaches zero as $u$ goes to zero.

Thus, we have
\begin{align*}
& n\min_{0 \le s \le 1}\left(C_W^{\dm}s + \frac{R_2}{\sqrt{n}} s+\psi_P(s)\right)
=
n \left(C_W^{\dm}s_n + \frac{R_2}{\sqrt{n}} s_n+\psi_P(s_n)\right)
=n (\frac{R_2}{\sqrt{n}} s_n 
+\int_{0}^{s_n}\int_{0}^{t} \frac{d^2 \psi_P}{ds^2}(u) du dt)\\
= &
\sqrt{n}R_2 s_n + n \frac{s_n^2}{2}\frac{d^2 \psi_P}{ds^2}(0)
+n \int_{0}^{s_n}\int_{0}^{t} \epsilon(u) du dt
\to 
\frac{-R_2^2 }{2\frac{d^2 \psi_P}{ds^2}(0)},
\end{align*}
which implies (\ref{6-16-1}).

\bibliographystyle{IEEE}

\end{document}